\documentclass[twocolumn,
superscriptaddress,
floatfix,
showkeys,apl,longbibliography]{revtex4-2}

\usepackage{graphicx} 
\usepackage{bm}	
\usepackage{color}                     
\usepackage{xcolor}
\usepackage{epsfig}
\usepackage{amsmath} 
\usepackage{amssymb} 
\usepackage[normalem]{ulem}
\usepackage{float}
\usepackage{mathtools}
\usepackage{xparse}
\usepackage{hyperref}
\usepackage{times}
\usepackage[normalem]{ulem}
\usepackage{textcomp}

\usepackage[caption=false]{subfig}

\usepackage[export]{adjustbox}
\usepackage{longtable}

\begin{document}
	
	\title{Generation of High-Resolution Handwritten Digits with an Ion-Trap Quantum Computer}
	
	\author{Manuel S. Rudolph}
	\affiliation{Zapata Computing Canada Inc., 325 Front St W, Toronto, ON, M5V 2Y1, Canada}
	
	\author{Ntwali Bashige Toussaint}
	\affiliation{Zapata Computing Inc., 100 Federal Street, Boston, MA 02110, USA}

	\author{Amara Katabarwa}
	\affiliation{Zapata Computing Inc., 100 Federal Street, Boston, MA 02110, USA}

	\author{Sonika Johri}
	\affiliation{IonQ Inc., 4505 Campus Drive, College Park, MD 20740, USA}
	
	\author{Borja Peropadre}
	\affiliation{Zapata Computing Inc., 100 Federal Street, Boston, MA 02110, USA}
	
	\author{Alejandro Perdomo-Ortiz}
	\email{alejandro@zapatacomputing.com}
	\affiliation{Zapata Computing Canada Inc., 325 Front St W, Toronto, ON, M5V 2Y1, Canada}
	
	
	\date{\today} 
	
	\begin{abstract}
		Generating high-quality data (e.g. images or video) is one of the most exciting and challenging frontiers in unsupervised machine learning. Utilizing quantum computers in such tasks to potentially enhance conventional machine learning algorithms has emerged as a promising application, but poses big challenges due to the limited number of qubits and the level of gate noise in available devices. In this work, we provide the first practical and experimental implementation of a quantum-classical generative algorithm capable of generating high-resolution images of handwritten digits with state-of-the-art gate-based quantum computers. In our quantum-assisted machine learning framework, we implement a quantum-circuit based generative model to learn and sample the prior distribution of a Generative Adversarial Network. We introduce a multi-basis technique which leverages the unique possibility of measuring quantum states in different bases, hence enhancing the expressivity of the prior distribution. We train this hybrid algorithm on an ion-trap device based on $^{171}$Yb$^{+}$ ion qubits to generate high-quality images and quantitatively outperform comparable classical Generative Adversarial Networks trained on the popular MNIST data set for handwritten digits.
	\end{abstract}
	
	\maketitle
	\section{Introduction}
	In the last decades, machine learning (ML) algorithms have significantly increased in importance and value due to the rapid progress in ML techniques and computational resources~\cite{LeCun-Nature-2015,SCHMIDHUBER2015deeplearning}. 
	However, even state-of-the-art algorithms face significant challenges in learning and generalizing from an ever increasing volume of unlabeled data~\cite{Cheng2018Neural,novak2018generalization,neyshabur2018overparametrization}. 
	With the advent of quantum computing, quantum algorithms for ML arise as natural candidates in the search of applications of noisy intermediate-scale quantum (NISQ) devices, with the potential to surpass classical ML capabilities~\cite{Preskill2018}. Among the top candidates to achieve a quantum advantage in ML are generative models~\cite{PerdomoOrtiz2017}, i.e. probabilistic models aiming to capture the most essential features of complex data and to generate similar data by sampling from the trained model distribution. 
	Although there has been promising progress towards demonstrating a \textit{quantum supremacy} for specific quantum computing tasks~\cite{Google2019supremacy,Pan2020Supremacy}, and quantum generative models have been proven to learn distributions which are outside of classical reach~\cite{du2018expressive,Glasser2019,sweke2020learnability, Coyle2019, hinsche2021learnability}, it is not clear whether these theoretical guarantees hold in practice, or whether such enhancements provided by a quantum generative model are limited to cases where one can prove a theoretical gap between classical and quantum algorithms. 
	
	In particular, quantum resources offer a divergent set of tools for tackling various challenges and could instead lead to a \textit{practical quantum advantage} by avoiding pitfalls of conventional classical algorithms, for example, by improving training and consequently enhancing performance on generative tasks.
	
	Despite all promises, applying and scaling quantum models on small quantum devices to tackle real-world data sets remains a big challenge for quantum ML algorithms. Ref.~\cite{PerdomoOrtiz2017} proposes to enable quantum models for practical application by exploiting the known dimensionality-reduction capabilities of deep neural networks~\cite{hinton2006dimensionality} and compressing classical data before it is handed to a small quantum device (see Ref.~\cite{qi2021qtnvqc} for a framework utilizing tensor networks for data compression). 
	Having a quantum model learn the so-called \textit{latent} representation of data and take part in a joint quantum-classical training loop, opens up hybrid models to leverage quantum resources and potentially enhance performance when compared to purely classical algorithms.
	This synergistic interaction between a quantum model and classical deep neural networks is at the heart of the proposed quantum-assisted Helmholtz machine~\cite{PerdomoOrtiz2017,Benedetti2017b} and more recent hybrid proposals~\cite{Wilson2019QAAN,Anschuetz2019QAAN} for enhancing Associative Adversarial Networks (AAN)~\cite{arici2016associative}.
	In the specific case of Ref.~\cite{Anschuetz2019QAAN}, the authors propose to use a Quantum Boltzmann Machine (QBM)~\cite{Amin2016}, while Refs.~\cite{Benedetti2017b,Wilson2019QAAN} experimentally demonstrated this concept with a D-Wave 2000Q annealing device. A similar adoption of this hybrid strategy with quantum annealers has been explored with variational autoencoders~\cite{Vinci2020VAE}. Despite these efforts, a definite demonstration utilizing truly quantum resources on NISQ devices and with full-size ML data sets, e.g. the MNIST data set of handwritten digits~\cite{LeCun1998MNIST}, has remained elusive to date. Recent experimental results on gate-based quantum computers~\cite{huang2020QGANmnist} illustrate that current proposals are far from generating high-quality MNIST digits.
	
	\begin{figure*}
		\centering
		\includegraphics[width= 0.84\textwidth]{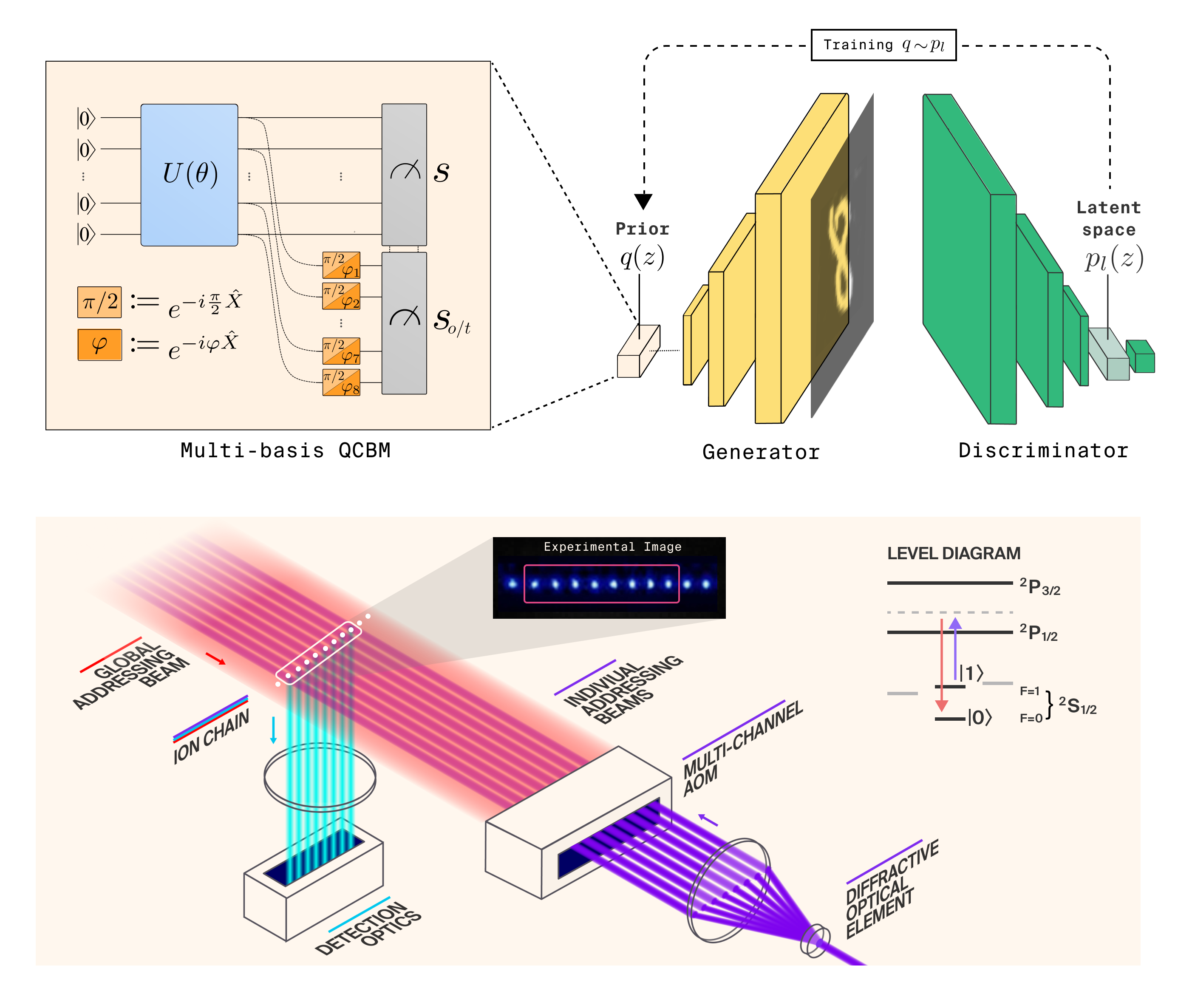}
		\caption{
			Top: Schematic description of our Quantum Circuit Associative Adversarial Network (QC-AAN) framework where the prior of a Generative Adversarial Network (GAN) is modelled by a multi-basis Quantum Circuit Born Machine (QCBM). This quantum generative model with encoded distribution $q(z)$ is trained on activations $z$ in the latent space of the discriminator, learning the feature distribution $p_l$. In the multi-basis QCBM, trainable single-qubit rotations follow the parametrized quantum circuit $U(\bm\theta)$ and allow for measuring the prepared wavefunction in additional bases. The angles for these post-rotations can be fixed, e.g. at $\pi/2$ to measure $s_o$ with all qubits in the \textit{orthogonal} $Y$-basis, or they can be \textit{trained} along with other parameters in $U(\bm\theta)$ to measure $s_t$. Measurements in computational basis $s$ and $s_{o/t}$ are concatenated and forwarded as prior samples into the generator network. The GAN is otherwise trained conventionally.
			Bottom: Illustration of the 11-qubit ion-trap quantum device by IonQ based on $^{171}$Yb$^{+}$ ion qubits. The experimental implementation of the QC-AAN algorithm in this work was performed on 8 qubits. The device is operated with automated loading of a linear chain of ions, which is then optically initialized with high fidelity. Computations are performed using a mode-locked $355$\,nm laser, which drives native single-qubit and two-qubit gates.}
		\label{fig:full_model}
	\end{figure*}
	In this work, we introduce the Quantum Circuit Associative Adversarial Network (QC-AAN): a framework combining capabilities of NISQ devices with classical deep learning techniques for generative modelling to learn relevant full-scale data sets (see Fig.~\ref{fig:full_model}).
	The framework applies a Quantum Circuit Born Machine (QCBM)~\cite{Benedetti2019} to model and re-parametrize the prior distribution of a Generative Adversarial Network (GAN)~\cite{goodfellow2014generative}. Furthermore, we introduce a multi-basis technique for the QCBM and argue that the use of a quantum generative model could enhance deep generative algorithms by providing them with non-classical distributions and quantum samples from a variety of measurement bases. Finally, to demonstrate the readiness of this framework, we train the QC-AAN with an experimental implementation of 8 qubits to generate the first high-resolution handwritten digits with end-to-end training of the quantum component on an ion-trap quantum device. Our results imply that near-term quantum devices could effectively be employed for generative modelling and flexibly assist classical GANs in their learning task.\\
	
	In Section~\ref{sec:GANs} we discuss GANs, which are the classical component of the QC-AAN, followed by the AAN framework in Section~\ref{sec:AANs}, which is aimed at solving common issues related to conventional GANs by providing them with an informed prior distribution. In Section~\ref{sec:QCBM} we introduce QCBMs and our multi-basis technique which can be used to possibly enhance the prior of a GAN with quantum measurements. Lastly, in Section~\ref{sec:experiments} we present simulation results and demonstrate the QC-AAN algorithm on hardware.
	
	\section{Generative Adversarial Networks (GAN\lowercase{s})}\label{sec:GANs}
	GANs~\cite{goodfellow2014generative} are one of the most popular recent generative machine learning algorithms able to generate remarkably realistic images and other data~\cite{radford2016dcgan,brock2019BigGAN}. The algorithm consists of a generator $G$ and a discriminator $D$, both of which are commonly implemented as deep artificial neural networks. The goal of a GAN is to train $G$ until its generated data is of satisfactory quality. This is achieved by using $D$ as a proxy for estimating the training loss and backpropagating the gradients to $G$. To that effect, $D$ performs a binary classification task to distinguish between data $x$ from the training dataset $p_{data}$ and data $\Tilde{x}$ which was generated data by $G$.
	The adversarial min-max loss function of a GAN can be written as
	\begin{equation}\label{eq:GAN_costfunction}
	\begin{aligned}
	\mathcal{C}_{GAN} = \min_G \max_D \Big[&\mathbb{E}_{x\sim p_{data}}[\log D(x)]\ +\\
	&\mathbb{E}_{\Tilde{x} \sim G}[\log\left(1-D(\Tilde{x})\right)]   \Big],
	\end{aligned}
	\end{equation}
	where $\mathbb{E}_{x \sim p_{data}}$ and $\mathbb{E}_{\Tilde{x} \sim G}$ denote expectations over data sampled from the training dataset $p_{data}$ and the generator $G$, respectively. 
	The outputs of $D$ are in the domain $D(x),D(\Tilde{x})\in(0,1)$, and it is trained to maximize $\mathcal{C}_{GAN}$ via $D(x)\xrightarrow{}1$ and $D(\Tilde{x})\xrightarrow{}0$. Conversely, $G$ is trained to minimize $\mathcal{C}_{GAN}$ by generating data which $D$ cannot classify with certainty ($D(x), D(\Tilde{x})\xrightarrow{}0.5$). While GANs have been shown to be able to generate remarkable data, common challenges in training a GAN lie in mode-collapse and non-convergence~\cite{goodfellow2014generative,brock2019BigGAN}, which are natural consequences of the delicately balanced adversarial game. One critical component of a GAN, which we argue is under-studied and we aim to enhance with a quantum model, is the source of randomness in $G$, i.e., the \textit{prior} distribution $q(z)$ in Fig.~\ref{fig:full_model}. $G$ takes prior samples $z$ as input to generate $\Tilde{x}=G(z)$. In other words, the neural network of $G$ learns a transformation between $q(z)$ and a high-quality data space. Because the model needs to be sampled efficiently on a classical computer, the $N$-dimensional prior is conventionally implemented as a continuous uniform distribution (i.e., $z\in (0,1)^{\otimes N}$) or a normal distribution (i.e., $z\in \mathcal{N}(\mu,\sigma)^{\otimes N}$ with mean $\mu=0$ and width $\sigma$). Discrete Bernoulli priors (i.e., $z\in \{0,1\}^{\otimes N}$) have also been shown empirically to be competitive~\cite{brock2019BigGAN} and are used throughout this work for the classical GANs. For such \textit{uninformed} prior choices, a prior with large $N$ requires a notably expressive neural network architecture to be able to map the full prior space to high-quality outputs, whereas a small $N$ could potentially lead to the algorithm not learning a good approximation of the full target data~\cite{padala2020prior}. Consequently, ML practitioners often rely on sufficiently large $N$ and scale the number of parameters in $G$ for their purpose. Clearly, an uninformed prior distribution is unlikely to be ideal for any given dataset and GAN architecture, and the prior distribution $q(z)$ should, if possible, be adapted such that $G$ can effectively map prior samples to a high-quality output space. The AAN framework~\cite{arici2016associative} aims to address all of these challenges by implementing a non-trivial prior distribution for $G$.
	
	\section{Associative Adversarial Networks (AAN\lowercase{s})}\label{sec:AANs}
	A schematic overview of the AAN framework can be viewed in Fig.~\ref{fig:full_model}. In an AAN, the prior is modelled by a smaller generative model with distribution $q_\theta(z)$ which adapts to the training data, the GAN architecture, and the current stage of training. The parameters $\theta$ of the prior are tuned to model the node activations in a deep layer $l$ of the discriminator $D$ with matching size $N$. This layer $l$ is called the \textit{latent space} layer and it captures highly condensed features of the training data and generated data (see Fig.~\ref{fig:full_model}). The prior is trained to maximize the likelihood of its parametrized distribution $q_\theta(z)$ given latent data samples $z$ from the latent data distribution $p_l(z)$:
	\begin{equation}\label{eq:cost_prior}
	\mathcal{C}_q = \max_q \mathbb{E}_{z\sim p_l(z)}\left[\log q_\theta(z)\right],
	\end{equation}
	where $\mathbb{E}_{z\sim p_l(z)}$ indicates the expectation over the observed latent data samples.
	By training $q_\theta(z)$ to maximize $\mathcal{C}_q$, $G$ receives explicit access to information which $D$ deems to be important for its classification task.
	
	Although the original AAN work proposed using Restricted Boltzmann Machines (RBMs)~\cite{ackley1985RBM} to model the prior, RBMs have been shown to be outperformed by comparable QCBMs in learning and sampling probability distributions constructed from real-world data~\cite{Alcazar2020ClvsQuant}. Therefore, in this work, we implement a QCBM in the prior of a GAN to tackle common challenges of GANs and potentially enhancing them with measurements from quantum distributions. As GANs typically employ 16 to 128 dimensional priors (see Appendix E in Ref.~\cite{brock2019BigGAN}), we expect our approach to be able to produce very competitive results on practical datasets by using near to mid-term quantum devices.
	
	\section{Quantum Circuit Born Machines \&\\The Multi-Basis Technique}\label{sec:QCBM}
	A QCBM is a circuit-based generative model which encodes a data distribution in a quantum state. This approach allows for sampling of the QCBM by repeatedly preparing and measuring its corresponding wavefunction
	\begin{equation}\label{eq:qcbm_wavefunction}
	|\psi(\bm\theta)\rangle= \text{U}(\bm\theta)|0\rangle.
	\end{equation}
	$\text{U}(\bm\theta)$ is a parametrized quantum circuit acting on an initial qubit state $|0\rangle$, with $\text{U}$ chosen according to the capabilities and limitations of NISQ devices. The probabilities for observing any of the $2^{n}$ bitstrings $\bm{s}$ in the $n$-bit (qubit) target probability distribution are modeled using the Born probabilities such that 
	\begin{equation}\label{eq:QCBM_born_probabilities}
	q_\theta(\bm{s})  =  | \langle \bm{s} |\psi(\bm\theta)\rangle |^2.
	\end{equation}
	Importantly, QCBMs can be implemented on most NISQ devices (see e.g. Refs.~\cite{Hamilton2018,leyton2019robust,coyle2020generativeFinance,Zhu2018, hamilton2020hardwareQCBM}) and additionally open our algorithm up to exploit unique features of quantum circuit-based approaches, like measuring in a range of bases.

	By training a QCBM on computational basis samples, families of sample distributions, i.e. projections of the wavefunction, become accessible in a range of other basis sets. This is information which is present in the QCBM but which is conventionally not used when measuring only in computational basis. Thus, we propose a multi-basis technique for the QCBM which provides the QC-AAN with a prior space consisting of quantum samples in flexible bases. The multi-basis technique can potentially enhance the performance of the generator by providing it with quantum samples in different measurements bases which have no classical analog. Follow-up work has indeed indicated that quantum measurements in different bases can lead to a separation of quantum and classical generative models~\cite{gao2021enhancing}. In principle, one could construct and measure many different basis sets, although this implies measuring increasingly redundant bases. The optimal number of basis sets likely depends on the given learning task and available quantum resources. In this work, we restrict ourselves to measurements in one additional basis set, as well as bases which can be reached by one layer of single-qubit rotations. This prevents adding significant depth to the QCBM circuit which is later implemented on quantum hardware. One could however measure in bases which are generated by more complex rotations.
	
	Fig.~\ref{fig:full_model} displays how the multi-basis technique is applied in practice. 
	For each measurement \textbf{s} in computational basis according to Eq.~\ref{eq:QCBM_born_probabilities}, a second measurement is prepared by applying parametrized single qubit rotations $R_X(\varphi_i)$ to the wavefunction for each qubit $i$ and with parameters $\varphi$. For $\varphi = \pi/2$, each qubit is rotated into the Y-basis, which we refer to as the `orthogonal basis' and is denoted with $o$ throughout this work. The more general case defines what we call the `trained basis' approach, which we denote with $t$. Here, the parameters $\varphi$ are trained for each qubit along with other circuit parameters to optimize the information extracted from the quantum state.  
	When preparing the QCBM wavefunction in Eq.~\ref{eq:qcbm_wavefunction} and applying the corresponding single-qubit post-rotations, we obtain a sample $\textbf{s}_{o/t}$ in the orthogonal or trained basis, respectively. Both measurements \textbf{s} and $\textbf{s}_{o/t}$ are then forwarded through a fully-connected neural network layer and into $G$ to learn an effective utilization of the provided quantum resources. These multi-basis variants of the QC-AAN are called QC$_{+o}$-AAN and QC$_{+t}$-AAN, respectively. Implementation and training details of the multi-basis QCBM, we refer to Appendix~\ref{Apx:QCBM_overview} and Appendix~\ref{Apx:qcbm_training}.

	\begin{figure}
		\centering
		\includegraphics[width=1\linewidth]{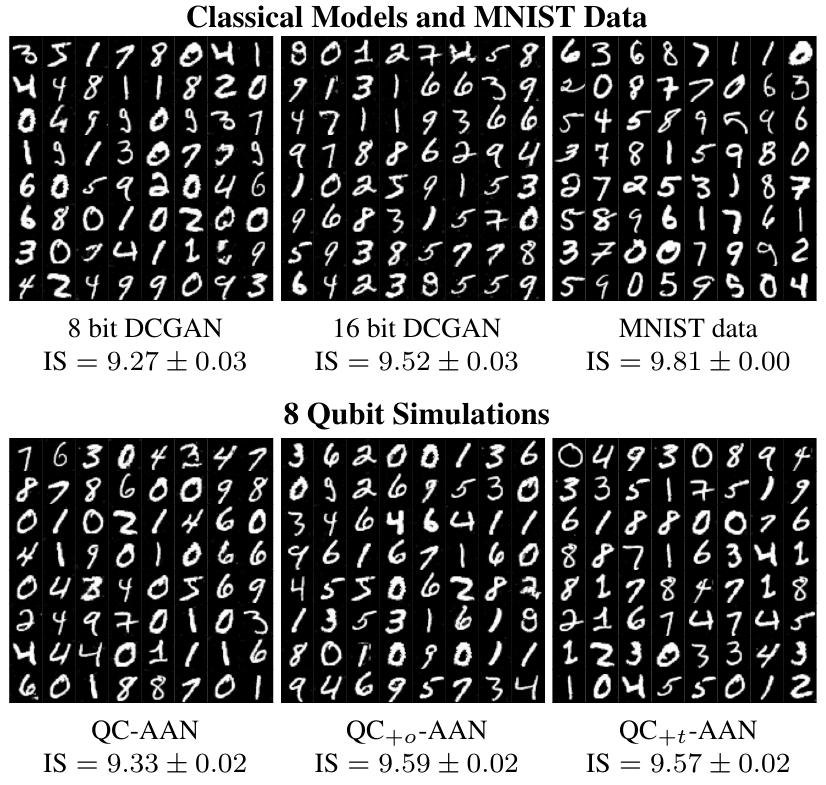}
		\caption{Digits generated by our QC-AANs with 8 simulated qubits and comparable classical DCGANs. The differences in performance are only due to replacing the uninformed random prior of the DCGANs with a multi-basis QCBM. All neural network architectures are equivalent. The models shown here were selected among several training repetitions to have a high Inception Score for their respective model type. The multi-basis technique in the QC$_{+o/t}$-AAN enhances the algorithm compared to the 8 and 16-bit DCGAN, and the ``plain vanilla" QC-AAN (i.e.,without measurements in a second basis.)}
		\label{fig:digits_8_qubits}\hfill
	\end{figure}
	\begin{figure*}
		\centering
		\includegraphics[width=0.95\linewidth]{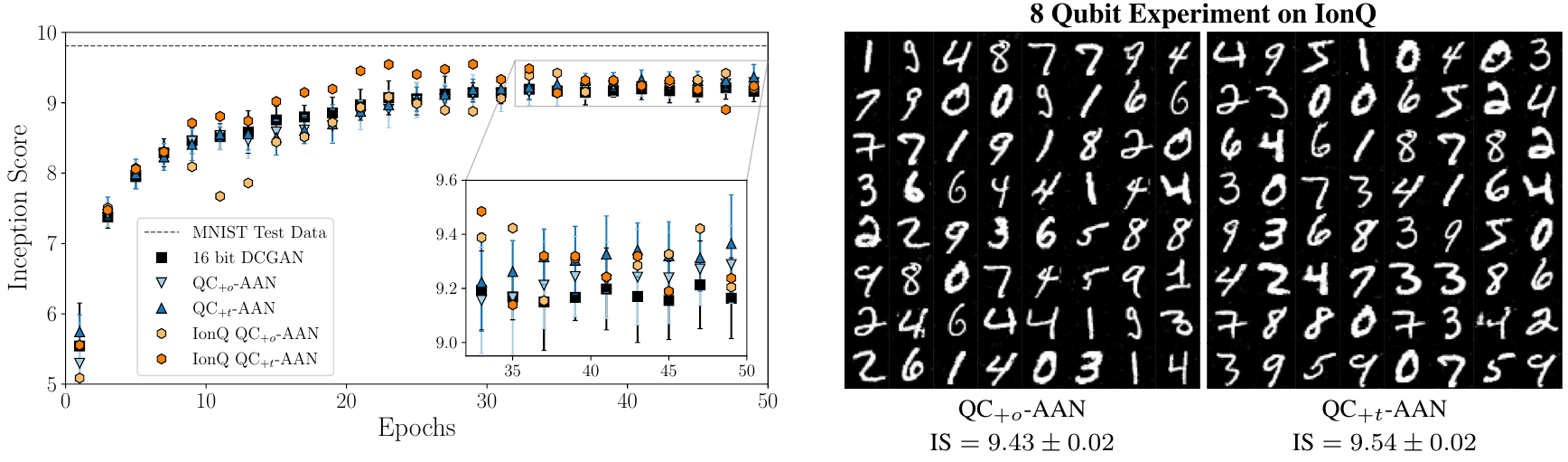}
		\caption{Left: Quantitative comparison between DCGANs with 16 bit random prior distribution and our 8 qubit QC$_{+o/t}$-AAN algorithm. The experimental realization on the IonQ device includes complete implementation of the multi-basis QCBM on hardware. Where present, error bars indicate the standard deviation of 10 independent training repetitions. The 8 qubit hybrid models generally outperform the classical DCGAN with uninformed prior distribution while the neural network architectures are equivalent.  Right: Images of handwritten digits generated by the experimental implementation of the QC$_{+o/t}$-AAN models on the IonQ device. The QC$_{+o}$-AAN model achieves a maximal Inception score of over 9.4 while the QC$_{+t}$-AAN scores over $9.5$ with overall better diversity.}
		\label{fig:IonQ_full_run_images}
	\end{figure*}
	\section{Applying the QC-AAN Framework}\label{sec:experiments}
	As a first step towards showcasing the QC-AAN and our multi-basis technique, we numerically simulate training on the canonical MNIST data set of handwritten digits~\cite{LeCun1998MNIST}, a standard data set for benchmarking a variety of ML and deep learning algorithms. All implementations in this work are performed using the Orquestra\texttrademark\ platform. To isolate the effect of modelling the prior with a QCBM, we compare our quantum-classical models to purely classical Deep Convolutional GANs (DCGANs) with precisely the same neural network architecture (see Appendix~\ref{Apx:network_achitecture} for details) and with uniform prior distribution. The QCBM prior in the QC-AAN adds between $31$ and $52$ trainable parameters, which equates to an increase of approximately $0.02\%$ in the total number of parameters relative to the reference DCGAN used in this work.
	
	The QCBM is initiated with a warm start such that the prior distribution is uniform and thus QC-AANs and DCGANs are equivalent at the beginning of training. This initialization provides us with a good baseline for comparison and should additionally avoid complications related to barren plateaus~\cite{Mcclear2018Barren}.
	We then employ a smooth transition training protocol where the QCBM slowly follows changes in the latent space of $D$ to facilitate stable training of $G$ (see Appendix~\ref{Apx:qcbm_training} for details). As shown in Appendix~\ref{Apx:rbm_qcbm}, the robust global sampling abilities of the QCBM have shown to be very useful in this training protocol where we have observed instabilities with the RBM. 
	To quantitatively assess performance, we calculate the Inception Score (IS) (see Appendix~\ref{Apx:IS}) which evaluates the quality and diversity of generated images in GANs. The IS is high for a model which produces diverse images of high-quality handwritten digits. However, even the MNIST dataset does not achieve a perfect IS of $10$, as some digits are not clearly identifiable.
	
	Fig.~\ref{fig:digits_8_qubits} shows results of handwritten digits generated by our models.
	For each model type, we pre-selected the best-performing models in terms of the IS and, among those, chose a single representative based on appearance of the images for a human observer. The generated digits themselves are random sub-samples of the selected models.
	It is apparent that all models presented here can achieve good performance and output high-resolution handwritten digits. This is generally expected as the MNIST dataset is considered to be a simple dataset to learn for modern network architectures. In a quantitative evaluation of average model performance (see also Appendix~\ref{Apx:simulated_results}), we find that the 8 qubit QC-AAN without multi-basis technique typically does not outperform comparable 8 bit DCGANs under any of the hyperparameters explored. For low-dimensional priors in general, we were not able to improve the performance of uniform prior distributions.
	In contrast to that, both multi-basis QC-AAN models, the QC$_{+o}$-AAN and the QC$_{+t}$-AAN, generate visibly better images and achieve higher scores than the 8 bit and 8 qubit models without additional basis samples. In fact, Fig.~\ref{fig:IonQ_full_run_images} shows that, with an average IS of $9.28$ and $9.36$, respectively, both multi-basis models outperform the 16 bit DCGAN with an average IS of $9.20$. Considering that the values are expected to saturate towards the IS of the training data ($\text{IS}=9.81$), these are remarkable results. Additionally, we observe that an 8 qubit multi-basis QCBM does not require full access to a 16 qubit Hilbert space to outperform a 16 bit random prior, and that the trained-basis approach generally enhances the algorithm even more compared to the fixed orthogonal-basis approach. We emphasize that these differences in performance were achieved merely by implementing a multi-basis QCBM as informed prior to the generator.
	
	To provide final confirmation that the QC-AAN framework is fit for implementation on NISQ devices, we train both QC$_{+o/t}$-AAN algorithms on an 11-qubit quantum device from IonQ which is based on $^{171}$Yb$^{+}$ ion qubits. For details on the device, we refer to Fig.~\ref{fig:full_model}, Appendix~\ref{Apx:IonQ_device}, and Ref.~\cite{Wright2019IonQ}. To achieve high fidelity and to attend to limitations of the quantum hardware, we consider the following adaptations as compared to the simulation setup. While this device allows for all-to-all connectivity, long-range gates are generally more noisy. We thus employ a next-neighbor entangling topology for in our quantum circuits. Additionally, we utilize parametrized M\o lmer S\o renson entangling gates (see Eq.~\ref{eq:ms_gate}) in the circuit ansatz of the QCBM which are native to the ion-trap device. This reduces gate overhead and thus noise introduced from gate re-compilation. Finally, to reduce the overall number of calls to the quantum device, as well as wall time between evaluations of the same quantum circuit, we aggregate a large number of measurements in-between training steps of the quantum prior. To optimize run time of the algorithm, one could simultaneously perform and buffer measurements on the quantum device while the classical networks are training with samples from a constant prior distribution. Only during training of the quantum circuit do all other components need to halt. For more information on the training protocol of the QCBM, we refer to Appendix~\ref{Apx:qcbm_training}.
	
	The experimental results for the training \textit{on hardware} can be viewed in Fig.~\ref{fig:IonQ_full_run_images}. Every image generated by the GAN during and after training has been produced exclusively utilizing hardware measurements from multiple bases. Note that we demonstrate the results of a single run on hardware per model type. No averaging or post-processing was performed.
	To the best of our knowledge, this is the first practical implementation of a quantum-classical algorithm capable of generating high-resolution digits on a NISQ device. 
	\section{Discussion \& Outlook}
	With as few as 8 qubits, we show signs of positively influencing the training of GANs and indicate general utility in modelling their prior with a multi-basis QCBM on NISQ devices. Learning the choice of the measurement bases through the quantum-classical training loop, i.e. our \mbox{QC$_{+t}$-AAN} algorithm, appears to be the most successful approach in simulations and also in the experimental realization on the IonQ device. This is a great example of how quantum components in a hybrid quantum ML algorithm are capable of effectively utilizing feedback coming from classical neural networks and a testament to the general ML approach of learning the best parameters rather than pre-determining them. 
	Unlike many other use-case implementations of quantum algorithms on NISQ devices, our models do not underperform compared to noise-free simulations. It is reasonable that significant re-parametrization of the prior space, paired with a modest noise floor, provides GANs with an improved trade-off between exploration of the target space and convergence to high-quality data.
	
	Our QC-AAN framework also extends flexibly to more complex data sets such as data with higher resolution and color, for which we expect refinement of the prior distribution to become more vital for performance of the algorithm. Besides extending to these more challenging data sets, we could adapt the learning strategy of the quantum prior to follow a different objective function as compared to the AAN framework, potentially one that directly ties into improving the generator's performance in the adversarial loss function in Eq.~\ref{eq:GAN_costfunction}. Because of the distributed nature of the quantum-classical optimization loop, one may additionally adapt the training to follow a secure learning protocol~\cite{sheng2017distributed, song2021secure, li2021blind} for the current scenario where quantum and classical components are owned by separate entities.
	
	Our work has been driven by the challenge of finding a practical generative learning task which is more successfully accomplished with a quantum generative model~\cite{PerdomoOrtiz2017}. This could for example be achieved by encoding classically out-of-reach quantum distributions~\cite{du2018expressive,Glasser2019,sweke2020learnability, Coyle2019, hinsche2021learnability}. In fact, recent work has used the idea introduced here of using measurements from multiple bases to prove one kind of such classical-quantum model separation~\cite{gao2021enhancing}. However, our work also emphasizes another form of practical quantum advantage which is usually dismissed, but arguably equally important towards reaching a high-performing generative model: the training trajectory in model space that needs to be traversed from the initial distribution to the target distribution. For example, we have observed an improvement in robustness of the QCBM prior as compared to the RBM, which we showcase in Appendix~\ref{Apx:rbm_qcbm}. Both quantum enhancement scenarios are schematically discussed and illustrated in Appendix~\ref{Apx:practical}. Consequently, it is essential that we better understand the capabilities of our multi-basis technique, which kind of quantum distributions can be built from families of basis measurements, and their practical impact on GAN training. Particularly, one needs to consider the trade-offs of the multi-basis technique against doubling the number of qubits in the QCBM. While the multi-basis model is less expressive, it has the advantage that it can be implemented on earlier quantum devices, has lower circuit depth, employs significantly less parameters, and may thus be more effectively trainable. Additionally, the multi-basis QCBM exhibits natural synergy with the subsequent neural network that transform the samples from different measurement bases. The seemingly rigid and structured bi-partition of the effective prior space is significantly mitigated by a fully-connected layer which has the potential to reorder and learn non-trivial correlations among the bits in the prior during training.
	
	Moving forward, the use of a quantum generative model like a QCBM as a building block in the prior space of a larger classical architecture could unlock the path to reliably investigating the impact of certain prior shapes during training of large generative ML models. This is a field which rather under-studied in the context of classical deep learning algorithms but has shown promising results~\cite{arici2016associative,brock2019BigGAN, Anschuetz2019QAAN, Wilson2019QAAN}. Given that they are highly flexible and can provide true global samples from the encoded distribution, quantum generative models may offer the tools required to enable this research more effectively. Implementing prior distributions from quantum-inspired models such as the tensor-network-based Born machines~\cite{Cheng2017TensorBorn} is another exciting research direction that we will be exploring. In terms of generative applications outside of image generation, we will be studying the QC-AAN as a generator in the quantum-enhanced framework for combinatorial optimization problems presented in Ref.~\cite{alcazar2021enhancing}. There, the QC-ANN allows quantum-enhanced optimizers to explore combinatorial problems with a larger number of variables than those considered with the quantum-inspired tensor network models, and it allows the framework to extend to non-discrete variables.
	
	Finally, a full-fledged quantitative comparison between quantum and classical versions of machine learning algorithms can be challenging. For instance, classical resources are currently much cheaper and more accessible than quantum resources. For the hardware experiments performed in this work, getting access to and training the quantum component takes significantly more time than sampling pseudo-random numbers for the conventional random GAN prior. Therefore, factors such as cost and training time, particularly for time-sensitive applications, need to be considered when deciding whether a model has reached practical quantum advantage. Our focus has been towards showcasing our new quantum-assisted algorithm which leverages quantum circuits and their unique capabilities of using measurements in multiple basis.
	Although for this task one might be able to achieve better classical ML performance by choosing more sophisticated neural networks or training techniques, our QC-AAN framework is adaptive to any GAN architecture and could potentially enhance even the best classical GANs with NISQ devices on full-scale data sets. With the recent metrics for evaluating the generalization capabilities of generative models~\cite{gili2022evaluating}, this is a question that can be studied in a grounded and quantitative manner.\\
	
	\begin{acknowledgments} 
		The authors would like to thank Coleman Collins and Algert Sula for their support with the experimental images and the design of the hardware illustrations. The authors also thank Yudong Cao, Max Radin, Marta Mauri, Matthew Beach, Dax Enshan Koh, and J\'{e}r\^{o}me F. Gonthier for their feedback on an early version of this manuscript. The authors would like to acknowledge Zapata Computing's Platform Team for all the support with Orquestra\texttrademark: the software platform used during the execution of all the simulations and  experiments shown here. M.S.R. would like to acknowledge Zapata Computing for hosting his Quantum Applications Internship.

	\end{acknowledgments}
	
	\twocolumngrid{}
	

\begin{thebibliography}{51}%
		\makeatletter
		\providecommand \@ifxundefined [1]{%
			\@ifx{#1\undefined}
		}%
		\providecommand \@ifnum [1]{%
			\ifnum #1\expandafter \@firstoftwo
			\else \expandafter \@secondoftwo
			\fi
		}%
		\providecommand \@ifx [1]{%
			\ifx #1\expandafter \@firstoftwo
			\else \expandafter \@secondoftwo
			\fi
		}%
		\providecommand \natexlab [1]{#1}%
		\providecommand \enquote  [1]{``#1''}%
		\providecommand \bibnamefont  [1]{#1}%
		\providecommand \bibfnamefont [1]{#1}%
		\providecommand \citenamefont [1]{#1}%
		\providecommand \href@noop [0]{\@secondoftwo}%
		\providecommand \href [0]{\begingroup \@sanitize@url \@href}%
		\providecommand \@href[1]{\@@startlink{#1}\@@href}%
		\providecommand \@@href[1]{\endgroup#1\@@endlink}%
		\providecommand \@sanitize@url [0]{\catcode `\\12\catcode `\$12\catcode
			`\&12\catcode `\#12\catcode `\^12\catcode `\_12\catcode `\%12\relax}%
		\providecommand \@@startlink[1]{}%
		\providecommand \@@endlink[0]{}%
		\providecommand \url  [0]{\begingroup\@sanitize@url \@url }%
		\providecommand \@url [1]{\endgroup\@href {#1}{\urlprefix }}%
		\providecommand \urlprefix  [0]{URL }%
		\providecommand \Eprint [0]{\href }%
		\providecommand \doibase [0]{https://doi.org/}%
		\providecommand \selectlanguage [0]{\@gobble}%
		\providecommand \bibinfo  [0]{\@secondoftwo}%
		\providecommand \bibfield  [0]{\@secondoftwo}%
		\providecommand \translation [1]{[#1]}%
		\providecommand \BibitemOpen [0]{}%
		\providecommand \bibitemStop [0]{}%
		\providecommand \bibitemNoStop [0]{.\EOS\space}%
		\providecommand \EOS [0]{\spacefactor3000\relax}%
		\providecommand \BibitemShut  [1]{\csname bibitem#1\endcsname}%
		\let\auto@bib@innerbib\@empty
		\bibitem [{\citenamefont {LeCun}\ \emph {et~al.}(2015)\citenamefont {LeCun},
			\citenamefont {Bengio},\ and\ \citenamefont {Hinton}}]{LeCun-Nature-2015}%
		\BibitemOpen
		\bibfield  {author} {\bibinfo {author} {\bibfnamefont {Y.}~\bibnamefont
				{LeCun}}, \bibinfo {author} {\bibfnamefont {Y.}~\bibnamefont {Bengio}},\ and\
			\bibinfo {author} {\bibfnamefont {G.}~\bibnamefont {Hinton}},\ }\bibfield
		{title} {\bibinfo {title} {{Deep Learning}},\ }\href
		{https://doi.org/10.1038/nature14539} {\bibfield  {journal} {\bibinfo
				{journal} {Nature}\ }\textbf {\bibinfo {volume} {521}},\ \bibinfo {pages}
			{436} (\bibinfo {year} {2015})}\BibitemShut {NoStop}%
		\bibitem [{\citenamefont {Schmidhuber}(2015)}]{SCHMIDHUBER2015deeplearning}%
		\BibitemOpen
		\bibfield  {author} {\bibinfo {author} {\bibfnamefont {J.}~\bibnamefont
				{Schmidhuber}},\ }\bibfield  {title} {\bibinfo {title} {Deep learning in
				neural networks: An overview},\ }\href@noop {} {\bibfield  {journal}
			{\bibinfo  {journal} {Neural Networks}\ }\textbf {\bibinfo {volume} {61}},\
			\bibinfo {pages} {85 } (\bibinfo {year} {2015})}\BibitemShut {NoStop}%
		\bibitem [{\citenamefont {{Cheng}}\ \emph {et~al.}(2018)\citenamefont
			{{Cheng}}, \citenamefont {{Wang}}, \citenamefont {{Zhou}},\ and\
			\citenamefont {{Zhang}}}]{Cheng2018Neural}%
		\BibitemOpen
		\bibfield  {author} {\bibinfo {author} {\bibfnamefont {Y.}~\bibnamefont
				{{Cheng}}}, \bibinfo {author} {\bibfnamefont {D.}~\bibnamefont {{Wang}}},
			\bibinfo {author} {\bibfnamefont {P.}~\bibnamefont {{Zhou}}},\ and\ \bibinfo
			{author} {\bibfnamefont {T.}~\bibnamefont {{Zhang}}},\ }\bibfield  {title}
		{\bibinfo {title} {Model compression and acceleration for deep neural
				networks: The principles, progress, and challenges},\ }\href@noop {}
		{\bibfield  {journal} {\bibinfo  {journal} {IEEE Signal Processing Magazine}\
			}\textbf {\bibinfo {volume} {35}},\ \bibinfo {pages} {126} (\bibinfo {year}
			{2018})}\BibitemShut {NoStop}%
		\bibitem [{\citenamefont {Novak}\ \emph {et~al.}(2018)\citenamefont {Novak},
			\citenamefont {Bahri}, \citenamefont {Abolafia}, \citenamefont {Pennington},\
			and\ \citenamefont {Sohl-Dickstein}}]{novak2018generalization}%
		\BibitemOpen
		\bibfield  {author} {\bibinfo {author} {\bibfnamefont {R.}~\bibnamefont
				{Novak}}, \bibinfo {author} {\bibfnamefont {Y.}~\bibnamefont {Bahri}},
			\bibinfo {author} {\bibfnamefont {D.~A.}\ \bibnamefont {Abolafia}}, \bibinfo
			{author} {\bibfnamefont {J.}~\bibnamefont {Pennington}},\ and\ \bibinfo
			{author} {\bibfnamefont {J.}~\bibnamefont {Sohl-Dickstein}},\ }\bibfield
		{title} {\bibinfo {title} {Sensitivity and generalization in neural networks:
				an empirical study},\ }\href@noop {} {\bibfield  {journal} {\bibinfo
				{journal} {arXiv:1802.08760}\ } (\bibinfo {year} {2018})}\BibitemShut
		{NoStop}%
		\bibitem [{\citenamefont {Neyshabur}\ \emph {et~al.}(2018)\citenamefont
			{Neyshabur}, \citenamefont {Li}, \citenamefont {Bhojanapalli}, \citenamefont
			{LeCun},\ and\ \citenamefont {Srebro}}]{neyshabur2018overparametrization}%
		\BibitemOpen
		\bibfield  {author} {\bibinfo {author} {\bibfnamefont {B.}~\bibnamefont
				{Neyshabur}}, \bibinfo {author} {\bibfnamefont {Z.}~\bibnamefont {Li}},
			\bibinfo {author} {\bibfnamefont {S.}~\bibnamefont {Bhojanapalli}}, \bibinfo
			{author} {\bibfnamefont {Y.}~\bibnamefont {LeCun}},\ and\ \bibinfo {author}
			{\bibfnamefont {N.}~\bibnamefont {Srebro}},\ }\bibfield  {title} {\bibinfo
			{title} {The role of over-parametrization in generalization of neural
				networks},\ }in\ \href@noop {} {\emph {\bibinfo {booktitle} {International
					Conference on Learning Representations}}}\ (\bibinfo {year}
		{2018})\BibitemShut {NoStop}%
		\bibitem [{\citenamefont {Preskill}(2018)}]{Preskill2018}%
		\BibitemOpen
		\bibfield  {author} {\bibinfo {author} {\bibfnamefont {J.}~\bibnamefont
				{Preskill}},\ }\bibfield  {title} {\bibinfo {title} {Quantum computing in the
				{NISQ} era and beyond},\ }\href {https://doi.org/10.22331/q-2018-08-06-79}
		{\bibfield  {journal} {\bibinfo  {journal} {Quantum}\ }\textbf {\bibinfo
				{volume} {2}},\ \bibinfo {pages} {79} (\bibinfo {year} {2018})}\BibitemShut
		{NoStop}%
		\bibitem [{\citenamefont {Perdomo-Ortiz}\ \emph {et~al.}(2018)\citenamefont
			{Perdomo-Ortiz}, \citenamefont {Benedetti}, \citenamefont
			{Realpe-G{\'o}mez},\ and\ \citenamefont {Biswas}}]{PerdomoOrtiz2017}%
		\BibitemOpen
		\bibfield  {author} {\bibinfo {author} {\bibfnamefont {A.}~\bibnamefont
				{Perdomo-Ortiz}}, \bibinfo {author} {\bibfnamefont {M.}~\bibnamefont
				{Benedetti}}, \bibinfo {author} {\bibfnamefont {J.}~\bibnamefont
				{Realpe-G{\'o}mez}},\ and\ \bibinfo {author} {\bibfnamefont {R.}~\bibnamefont
				{Biswas}},\ }\bibfield  {title} {\bibinfo {title} {Opportunities and
				challenges for quantum-assisted machine learning in near-term quantum
				computers},\ }\href {https://doi.org/10.1088/2058-9565/aab859} {\bibfield
			{journal} {\bibinfo  {journal} {Quantum Science and Technology}\ }\textbf
			{\bibinfo {volume} {3}},\ \bibinfo {pages} {030502} (\bibinfo {year}
			{2018})}\BibitemShut {NoStop}%
		\bibitem [{\citenamefont {Arute}\ \emph {et~al.}(2019)\citenamefont {Arute},
			\citenamefont {Arya}, \citenamefont {Babbush}, \citenamefont {Bacon},
			\citenamefont {Bardin}, \citenamefont {Barends}, \citenamefont {Biswas},
			\citenamefont {Boixo}, \citenamefont {Brandao}, \citenamefont {Buell} \emph
			{et~al.}}]{Google2019supremacy}%
		\BibitemOpen
		\bibfield  {author} {\bibinfo {author} {\bibfnamefont {F.}~\bibnamefont
				{Arute}}, \bibinfo {author} {\bibfnamefont {K.}~\bibnamefont {Arya}},
			\bibinfo {author} {\bibfnamefont {R.}~\bibnamefont {Babbush}}, \bibinfo
			{author} {\bibfnamefont {D.}~\bibnamefont {Bacon}}, \bibinfo {author}
			{\bibfnamefont {J.~C.}\ \bibnamefont {Bardin}}, \bibinfo {author}
			{\bibfnamefont {R.}~\bibnamefont {Barends}}, \bibinfo {author} {\bibfnamefont
				{R.}~\bibnamefont {Biswas}}, \bibinfo {author} {\bibfnamefont
				{S.}~\bibnamefont {Boixo}}, \bibinfo {author} {\bibfnamefont {F.~G.}\
				\bibnamefont {Brandao}}, \bibinfo {author} {\bibfnamefont {D.~A.}\
				\bibnamefont {Buell}}, \emph {et~al.},\ }\bibfield  {title} {\bibinfo {title}
			{Quantum supremacy using a programmable superconducting processor},\
		}\href@noop {} {\bibfield  {journal} {\bibinfo  {journal} {Nature}\ }\textbf
			{\bibinfo {volume} {574}},\ \bibinfo {pages} {505} (\bibinfo {year}
			{2019})}\BibitemShut {NoStop}%
		\bibitem [{\citenamefont {Zhong}\ \emph {et~al.}(2020)\citenamefont {Zhong},
			\citenamefont {Wang}, \citenamefont {Deng}, \citenamefont {Chen},
			\citenamefont {Peng}, \citenamefont {Luo}, \citenamefont {Qin}, \citenamefont
			{Wu}, \citenamefont {Ding}, \citenamefont {Hu} \emph
			{et~al.}}]{Pan2020Supremacy}%
		\BibitemOpen
		\bibfield  {author} {\bibinfo {author} {\bibfnamefont {H.-S.}\ \bibnamefont
				{Zhong}}, \bibinfo {author} {\bibfnamefont {H.}~\bibnamefont {Wang}},
			\bibinfo {author} {\bibfnamefont {Y.-H.}\ \bibnamefont {Deng}}, \bibinfo
			{author} {\bibfnamefont {M.-C.}\ \bibnamefont {Chen}}, \bibinfo {author}
			{\bibfnamefont {L.-C.}\ \bibnamefont {Peng}}, \bibinfo {author}
			{\bibfnamefont {Y.-H.}\ \bibnamefont {Luo}}, \bibinfo {author} {\bibfnamefont
				{J.}~\bibnamefont {Qin}}, \bibinfo {author} {\bibfnamefont {D.}~\bibnamefont
				{Wu}}, \bibinfo {author} {\bibfnamefont {X.}~\bibnamefont {Ding}}, \bibinfo
			{author} {\bibfnamefont {Y.}~\bibnamefont {Hu}}, \emph {et~al.},\ }\bibfield
		{title} {\bibinfo {title} {Quantum computational advantage using photons},\
		}\href@noop {} {\bibfield  {journal} {\bibinfo  {journal} {Science}\ }\textbf
			{\bibinfo {volume} {370}},\ \bibinfo {pages} {1460} (\bibinfo {year}
			{2020})}\BibitemShut {NoStop}%
		\bibitem [{\citenamefont {Du}\ \emph {et~al.}(2018)\citenamefont {Du},
			\citenamefont {Hsieh}, \citenamefont {Liu},\ and\ \citenamefont
			{Tao}}]{du2018expressive}%
		\BibitemOpen
		\bibfield  {author} {\bibinfo {author} {\bibfnamefont {Y.}~\bibnamefont
				{Du}}, \bibinfo {author} {\bibfnamefont {M.-H.}\ \bibnamefont {Hsieh}},
			\bibinfo {author} {\bibfnamefont {T.}~\bibnamefont {Liu}},\ and\ \bibinfo
			{author} {\bibfnamefont {D.}~\bibnamefont {Tao}},\ }\bibfield  {title}
		{\bibinfo {title} {Expressive power of parametrized quantum circuits},\
		}\href {https://doi.org/10.1103/PhysRevResearch.2.033125} {\bibfield
			{journal} {\bibinfo  {journal} {Physical Review Research}\ }\textbf {\bibinfo
				{volume} {2}},\ \bibinfo {pages} {033125} (\bibinfo {year}
			{2018})}\BibitemShut {NoStop}%
		\bibitem [{\citenamefont {Glasser}\ \emph {et~al.}(2019)\citenamefont
			{Glasser}, \citenamefont {Sweke}, \citenamefont {Pancotti}, \citenamefont
			{Eisert},\ and\ \citenamefont {Cirac}}]{Glasser2019}%
		\BibitemOpen
		\bibfield  {author} {\bibinfo {author} {\bibfnamefont {I.}~\bibnamefont
				{Glasser}}, \bibinfo {author} {\bibfnamefont {R.}~\bibnamefont {Sweke}},
			\bibinfo {author} {\bibfnamefont {N.}~\bibnamefont {Pancotti}}, \bibinfo
			{author} {\bibfnamefont {J.}~\bibnamefont {Eisert}},\ and\ \bibinfo {author}
			{\bibfnamefont {I.}~\bibnamefont {Cirac}},\ }\bibfield  {title} {\bibinfo
			{title} {Expressive power of tensor-network factorizations for probabilistic
				modeling},\ }\href@noop {} {\bibfield  {journal} {\bibinfo  {journal}
				{Advances in neural information processing systems}\ }\textbf {\bibinfo
				{volume} {32}} (\bibinfo {year} {2019})}\BibitemShut {NoStop}%
		\bibitem [{\citenamefont {Sweke}\ \emph {et~al.}(2021)\citenamefont {Sweke},
			\citenamefont {Seifert}, \citenamefont {Hangleiter},\ and\ \citenamefont
			{Eisert}}]{sweke2020learnability}%
		\BibitemOpen
		\bibfield  {author} {\bibinfo {author} {\bibfnamefont {R.}~\bibnamefont
				{Sweke}}, \bibinfo {author} {\bibfnamefont {J.-P.}\ \bibnamefont {Seifert}},
			\bibinfo {author} {\bibfnamefont {D.}~\bibnamefont {Hangleiter}},\ and\
			\bibinfo {author} {\bibfnamefont {J.}~\bibnamefont {Eisert}},\ }\bibfield
		{title} {\bibinfo {title} {On the quantum versus classical learnability of
				discrete distributions},\ }\href {https://doi.org/10.22331/q-2021-03-23-417}
		{\bibfield  {journal} {\bibinfo  {journal} {Quantum}\ }\textbf {\bibinfo
				{volume} {5}},\ \bibinfo {pages} {417} (\bibinfo {year} {2021})}\BibitemShut
		{NoStop}%
		\bibitem [{\citenamefont {Coyle}\ \emph {et~al.}(2020)\citenamefont {Coyle},
			\citenamefont {Mills}, \citenamefont {Danos},\ and\ \citenamefont
			{Kashefi}}]{Coyle2019}%
		\BibitemOpen
		\bibfield  {author} {\bibinfo {author} {\bibfnamefont {B.}~\bibnamefont
				{Coyle}}, \bibinfo {author} {\bibfnamefont {D.}~\bibnamefont {Mills}},
			\bibinfo {author} {\bibfnamefont {V.}~\bibnamefont {Danos}},\ and\ \bibinfo
			{author} {\bibfnamefont {E.}~\bibnamefont {Kashefi}},\ }\bibfield  {title}
		{\bibinfo {title} {The born supremacy: quantum advantage and training of an
				ising born machine},\ }\href@noop {} {\bibfield  {journal} {\bibinfo
				{journal} {npj Quantum Information}\ }\textbf {\bibinfo {volume} {6}}
			(\bibinfo {year} {2020})}\BibitemShut {NoStop}%
		\bibitem [{\citenamefont {Hinsche}\ \emph {et~al.}(2021)\citenamefont
			{Hinsche}, \citenamefont {Ioannou}, \citenamefont {Nietner}, \citenamefont
			{Haferkamp}, \citenamefont {Quek}, \citenamefont {Hangleiter}, \citenamefont
			{Seifert}, \citenamefont {Eisert},\ and\ \citenamefont
			{Sweke}}]{hinsche2021learnability}%
		\BibitemOpen
		\bibfield  {author} {\bibinfo {author} {\bibfnamefont {M.}~\bibnamefont
				{Hinsche}}, \bibinfo {author} {\bibfnamefont {M.}~\bibnamefont {Ioannou}},
			\bibinfo {author} {\bibfnamefont {A.}~\bibnamefont {Nietner}}, \bibinfo
			{author} {\bibfnamefont {J.}~\bibnamefont {Haferkamp}}, \bibinfo {author}
			{\bibfnamefont {Y.}~\bibnamefont {Quek}}, \bibinfo {author} {\bibfnamefont
				{D.}~\bibnamefont {Hangleiter}}, \bibinfo {author} {\bibfnamefont {J.-P.}\
				\bibnamefont {Seifert}}, \bibinfo {author} {\bibfnamefont {J.}~\bibnamefont
				{Eisert}},\ and\ \bibinfo {author} {\bibfnamefont {R.}~\bibnamefont
				{Sweke}},\ }\href@noop {} {\bibinfo {title} {Learnability of the output
				distributions of local quantum circuits}} (\bibinfo {year} {2021}),\ \Eprint
		{https://arxiv.org/abs/2110.05517} {arXiv:2110.05517 [quant-ph]} \BibitemShut
		{NoStop}%
		\bibitem [{\citenamefont {Hinton}\ and\ \citenamefont
			{Salakhutdinov}(2006)}]{hinton2006dimensionality}%
		\BibitemOpen
		\bibfield  {author} {\bibinfo {author} {\bibfnamefont {G.~E.}\ \bibnamefont
				{Hinton}}\ and\ \bibinfo {author} {\bibfnamefont {R.~R.}\ \bibnamefont
				{Salakhutdinov}},\ }\bibfield  {title} {\bibinfo {title} {Reducing the
				dimensionality of data with neural networks},\ }\href@noop {} {\bibfield
			{journal} {\bibinfo  {journal} {Science}\ }\textbf {\bibinfo {volume}
				{313}},\ \bibinfo {pages} {504} (\bibinfo {year} {2006})}\BibitemShut
		{NoStop}%
		\bibitem [{\citenamefont {Qi}\ \emph {et~al.}(2021)\citenamefont {Qi},
			\citenamefont {Yang},\ and\ \citenamefont {Chen}}]{qi2021qtnvqc}%
		\BibitemOpen
		\bibfield  {author} {\bibinfo {author} {\bibfnamefont {J.}~\bibnamefont
				{Qi}}, \bibinfo {author} {\bibfnamefont {C.-H.~H.}\ \bibnamefont {Yang}},\
			and\ \bibinfo {author} {\bibfnamefont {P.-Y.}\ \bibnamefont {Chen}},\
		}\bibfield  {title} {\bibinfo {title} {Qtn-vqc: An end-to-end learning
				framework for quantum neural networks},\ }\href@noop {} {\bibfield  {journal}
			{\bibinfo  {journal} {arXiv preprint arXiv:2110.03861}\ } (\bibinfo {year}
			{2021})}\BibitemShut {NoStop}%
		\bibitem [{\citenamefont {Benedetti}\ \emph
			{et~al.}(2018{\natexlab{a}})\citenamefont {Benedetti}, \citenamefont
			{Realpe-G{\'o}mez},\ and\ \citenamefont {Perdomo-Ortiz}}]{Benedetti2017b}%
		\BibitemOpen
		\bibfield  {author} {\bibinfo {author} {\bibfnamefont {M.}~\bibnamefont
				{Benedetti}}, \bibinfo {author} {\bibfnamefont {J.}~\bibnamefont
				{Realpe-G{\'o}mez}},\ and\ \bibinfo {author} {\bibfnamefont {A.}~\bibnamefont
				{Perdomo-Ortiz}},\ }\bibfield  {title} {\bibinfo {title} {Quantum-assisted
				helmholtz machines: A quantum--classical deep learning framework for
				industrial datasets in near-term devices},\ }\href
		{https://doi.org/10.1088/2058-9565/aabd98} {\bibfield  {journal} {\bibinfo
				{journal} {Quantum Science and Technology}\ }\textbf {\bibinfo {volume}
				{3}},\ \bibinfo {pages} {034007} (\bibinfo {year}
			{2018}{\natexlab{a}})}\BibitemShut {NoStop}%
		\bibitem [{\citenamefont {Wilson}\ \emph {et~al.}(2021)\citenamefont {Wilson},
			\citenamefont {Vandal}, \citenamefont {Hogg},\ and\ \citenamefont
			{Rieffel}}]{Wilson2019QAAN}%
		\BibitemOpen
		\bibfield  {author} {\bibinfo {author} {\bibfnamefont {M.}~\bibnamefont
				{Wilson}}, \bibinfo {author} {\bibfnamefont {T.}~\bibnamefont {Vandal}},
			\bibinfo {author} {\bibfnamefont {T.}~\bibnamefont {Hogg}},\ and\ \bibinfo
			{author} {\bibfnamefont {E.~G.}\ \bibnamefont {Rieffel}},\ }\bibfield
		{title} {\bibinfo {title} {Quantum-assisted associative adversarial network:
				Applying quantum annealing in deep learning},\ }\href@noop {} {\bibfield
			{journal} {\bibinfo  {journal} {Quantum Machine Intelligence}\ }\textbf
			{\bibinfo {volume} {3}},\ \bibinfo {pages} {1} (\bibinfo {year}
			{2021})}\BibitemShut {NoStop}%
		\bibitem [{\citenamefont {Anschuetz}\ and\ \citenamefont
			{Zanoci}(2019)}]{Anschuetz2019QAAN}%
		\BibitemOpen
		\bibfield  {author} {\bibinfo {author} {\bibfnamefont {E.~R.}\ \bibnamefont
				{Anschuetz}}\ and\ \bibinfo {author} {\bibfnamefont {C.}~\bibnamefont
				{Zanoci}},\ }\bibfield  {title} {\bibinfo {title} {Near-term
				quantum-classical associative adversarial networks},\ }\href
		{https://doi.org/10.1103/PhysRevA.100.052327} {\bibfield  {journal} {\bibinfo
				{journal} {Physical Review A}\ }\textbf {\bibinfo {volume} {100}},\ \bibinfo
			{pages} {052327} (\bibinfo {year} {2019})}\BibitemShut {NoStop}%
		\bibitem [{\citenamefont {Arici}\ and\ \citenamefont
			{Celikyilmaz}(2016)}]{arici2016associative}%
		\BibitemOpen
		\bibfield  {author} {\bibinfo {author} {\bibfnamefont {T.}~\bibnamefont
				{Arici}}\ and\ \bibinfo {author} {\bibfnamefont {A.}~\bibnamefont
				{Celikyilmaz}},\ }\bibfield  {title} {\bibinfo {title} {Associative
				adversarial networks},\ }\href@noop {} {\bibfield  {journal} {\bibinfo
				{journal} {arXiv:1611.06953}\ } (\bibinfo {year} {2016})}\BibitemShut
		{NoStop}%
		\bibitem [{\citenamefont {Amin}\ \emph {et~al.}(2018)\citenamefont {Amin},
			\citenamefont {Andriyash}, \citenamefont {Rolfe}, \citenamefont
			{Kulchytskyy},\ and\ \citenamefont {Melko}}]{Amin2016}%
		\BibitemOpen
		\bibfield  {author} {\bibinfo {author} {\bibfnamefont {M.~H.}\ \bibnamefont
				{Amin}}, \bibinfo {author} {\bibfnamefont {E.}~\bibnamefont {Andriyash}},
			\bibinfo {author} {\bibfnamefont {J.}~\bibnamefont {Rolfe}}, \bibinfo
			{author} {\bibfnamefont {B.}~\bibnamefont {Kulchytskyy}},\ and\ \bibinfo
			{author} {\bibfnamefont {R.}~\bibnamefont {Melko}},\ }\bibfield  {title}
		{\bibinfo {title} {Quantum {Boltzmann} machine},\ }\href@noop {} {\bibfield
			{journal} {\bibinfo  {journal} {Physical Review X}\ }\textbf {\bibinfo
				{volume} {8}} (\bibinfo {year} {2018})}\BibitemShut {NoStop}%
		\bibitem [{\citenamefont {Winci}\ \emph {et~al.}(2020)\citenamefont {Winci},
			\citenamefont {Buffoni}, \citenamefont {Sadeghi}, \citenamefont {Khoshaman},
			\citenamefont {Andriyash},\ and\ \citenamefont {Amin}}]{Vinci2020VAE}%
		\BibitemOpen
		\bibfield  {author} {\bibinfo {author} {\bibfnamefont {W.}~\bibnamefont
				{Winci}}, \bibinfo {author} {\bibfnamefont {L.}~\bibnamefont {Buffoni}},
			\bibinfo {author} {\bibfnamefont {H.}~\bibnamefont {Sadeghi}}, \bibinfo
			{author} {\bibfnamefont {A.}~\bibnamefont {Khoshaman}}, \bibinfo {author}
			{\bibfnamefont {E.}~\bibnamefont {Andriyash}},\ and\ \bibinfo {author}
			{\bibfnamefont {M.~H.}\ \bibnamefont {Amin}},\ }\bibfield  {title} {\bibinfo
			{title} {A path towards quantum advantage in training deep generative models
				with quantum annealers},\ }\href@noop {} {\bibfield  {journal} {\bibinfo
				{journal} {Machine Learning: Science and Technology}\ }\textbf {\bibinfo
				{volume} {1}},\ \bibinfo {pages} {045028} (\bibinfo {year}
			{2020})}\BibitemShut {NoStop}%
		\bibitem [{\citenamefont {{Lecun}}\ \emph {et~al.}(1998)\citenamefont
			{{Lecun}}, \citenamefont {{Bottou}}, \citenamefont {{Bengio}},\ and\
			\citenamefont {{Haffner}}}]{LeCun1998MNIST}%
		\BibitemOpen
		\bibfield  {author} {\bibinfo {author} {\bibfnamefont {Y.}~\bibnamefont
				{{Lecun}}}, \bibinfo {author} {\bibfnamefont {L.}~\bibnamefont {{Bottou}}},
			\bibinfo {author} {\bibfnamefont {Y.}~\bibnamefont {{Bengio}}},\ and\
			\bibinfo {author} {\bibfnamefont {P.}~\bibnamefont {{Haffner}}},\ }\bibfield
		{title} {\bibinfo {title} {Gradient-based learning applied to document
				recognition},\ }\href@noop {} {\bibfield  {journal} {\bibinfo  {journal}
				{Proceedings of the IEEE}\ }\textbf {\bibinfo {volume} {86}},\ \bibinfo
			{pages} {2278} (\bibinfo {year} {1998})}\BibitemShut {NoStop}%
		\bibitem [{\citenamefont {Huang}\ \emph {et~al.}(2021)\citenamefont {Huang},
			\citenamefont {Du}, \citenamefont {Gong}, \citenamefont {Zhao}, \citenamefont
			{Wu}, \citenamefont {Wang}, \citenamefont {Li}, \citenamefont {Liang},
			\citenamefont {Lin}, \citenamefont {Xu} \emph {et~al.}}]{huang2020QGANmnist}%
		\BibitemOpen
		\bibfield  {author} {\bibinfo {author} {\bibfnamefont {H.-L.}\ \bibnamefont
				{Huang}}, \bibinfo {author} {\bibfnamefont {Y.}~\bibnamefont {Du}}, \bibinfo
			{author} {\bibfnamefont {M.}~\bibnamefont {Gong}}, \bibinfo {author}
			{\bibfnamefont {Y.}~\bibnamefont {Zhao}}, \bibinfo {author} {\bibfnamefont
				{Y.}~\bibnamefont {Wu}}, \bibinfo {author} {\bibfnamefont {C.}~\bibnamefont
				{Wang}}, \bibinfo {author} {\bibfnamefont {S.}~\bibnamefont {Li}}, \bibinfo
			{author} {\bibfnamefont {F.}~\bibnamefont {Liang}}, \bibinfo {author}
			{\bibfnamefont {J.}~\bibnamefont {Lin}}, \bibinfo {author} {\bibfnamefont
				{Y.}~\bibnamefont {Xu}}, \emph {et~al.},\ }\bibfield  {title} {\bibinfo
			{title} {Experimental quantum generative adversarial networks for image
				generation},\ }\href@noop {} {\bibfield  {journal} {\bibinfo  {journal}
				{Physical Review Applied}\ }\textbf {\bibinfo {volume} {16}},\ \bibinfo
			{pages} {024051} (\bibinfo {year} {2021})}\BibitemShut {NoStop}%
		\bibitem [{\citenamefont {Benedetti}\ \emph
			{et~al.}(2018{\natexlab{b}})\citenamefont {Benedetti}, \citenamefont
			{Garcia-Pintos}, \citenamefont {Nam},\ and\ \citenamefont
			{Perdomo-Ortiz}}]{Benedetti2019}%
		\BibitemOpen
		\bibfield  {author} {\bibinfo {author} {\bibfnamefont {M.}~\bibnamefont
				{Benedetti}}, \bibinfo {author} {\bibfnamefont {D.}~\bibnamefont
				{Garcia-Pintos}}, \bibinfo {author} {\bibfnamefont {Y.}~\bibnamefont {Nam}},\
			and\ \bibinfo {author} {\bibfnamefont {A.}~\bibnamefont {Perdomo-Ortiz}},\
		}\bibfield  {title} {\bibinfo {title} {A generative modeling approach for
				benchmarking and training shallow quantum circuits},\ }\href
		{https://doi.org/10.1038/s41534-019-0157-8} {\bibfield  {journal} {\bibinfo
				{journal} {npj Quantum Information}\ }\textbf {\bibinfo {volume} {5}}
			(\bibinfo {year} {2018}{\natexlab{b}})}\BibitemShut {NoStop}%
		\bibitem [{\citenamefont {Goodfellow}\ \emph {et~al.}(2014)\citenamefont
			{Goodfellow}, \citenamefont {Pouget-Abadie}, \citenamefont {Mirza},
			\citenamefont {Xu}, \citenamefont {Warde-Farley}, \citenamefont {Ozair},
			\citenamefont {Courville},\ and\ \citenamefont
			{Bengio}}]{goodfellow2014generative}%
		\BibitemOpen
		\bibfield  {author} {\bibinfo {author} {\bibfnamefont {I.}~\bibnamefont
				{Goodfellow}}, \bibinfo {author} {\bibfnamefont {J.}~\bibnamefont
				{Pouget-Abadie}}, \bibinfo {author} {\bibfnamefont {M.}~\bibnamefont
				{Mirza}}, \bibinfo {author} {\bibfnamefont {B.}~\bibnamefont {Xu}}, \bibinfo
			{author} {\bibfnamefont {D.}~\bibnamefont {Warde-Farley}}, \bibinfo {author}
			{\bibfnamefont {S.}~\bibnamefont {Ozair}}, \bibinfo {author} {\bibfnamefont
				{A.}~\bibnamefont {Courville}},\ and\ \bibinfo {author} {\bibfnamefont
				{Y.}~\bibnamefont {Bengio}},\ }\bibfield  {title} {\bibinfo {title}
			{Generative adversarial nets},\ }\href@noop {} {\bibfield  {journal}
			{\bibinfo  {journal} {Advances in neural information processing systems}\
			}\textbf {\bibinfo {volume} {27}} (\bibinfo {year} {2014})}\BibitemShut
		{NoStop}%
		\bibitem [{\citenamefont {Radford}\ \emph {et~al.}(2015)\citenamefont
			{Radford}, \citenamefont {Metz},\ and\ \citenamefont
			{Chintala}}]{radford2016dcgan}%
		\BibitemOpen
		\bibfield  {author} {\bibinfo {author} {\bibfnamefont {A.}~\bibnamefont
				{Radford}}, \bibinfo {author} {\bibfnamefont {L.}~\bibnamefont {Metz}},\ and\
			\bibinfo {author} {\bibfnamefont {S.}~\bibnamefont {Chintala}},\ }\bibfield
		{title} {\bibinfo {title} {Unsupervised representation learning with deep
				convolutional generative adversarial networks},\ }\href@noop {} {\bibfield
			{journal} {\bibinfo  {journal} {arXiv:1511.06434}\ } (\bibinfo {year}
			{2015})}\BibitemShut {NoStop}%
		\bibitem [{\citenamefont {Brock}\ \emph {et~al.}(2018)\citenamefont {Brock},
			\citenamefont {Donahue},\ and\ \citenamefont {Simonyan}}]{brock2019BigGAN}%
		\BibitemOpen
		\bibfield  {author} {\bibinfo {author} {\bibfnamefont {A.}~\bibnamefont
				{Brock}}, \bibinfo {author} {\bibfnamefont {J.}~\bibnamefont {Donahue}},\
			and\ \bibinfo {author} {\bibfnamefont {K.}~\bibnamefont {Simonyan}},\
		}\bibfield  {title} {\bibinfo {title} {Large scale {GAN} training for high
				fidelity natural image synthesis},\ }\href@noop {} {\bibfield  {journal}
			{\bibinfo  {journal} {arXiv:1809.11096}\ } (\bibinfo {year}
			{2018})}\BibitemShut {NoStop}%
		\bibitem [{\citenamefont {Padala}\ \emph {et~al.}(2020)\citenamefont {Padala},
			\citenamefont {Das},\ and\ \citenamefont {Gujar}}]{padala2020prior}%
		\BibitemOpen
		\bibfield  {author} {\bibinfo {author} {\bibfnamefont {M.}~\bibnamefont
				{Padala}}, \bibinfo {author} {\bibfnamefont {D.}~\bibnamefont {Das}},\ and\
			\bibinfo {author} {\bibfnamefont {S.}~\bibnamefont {Gujar}},\ }\bibfield
		{title} {\bibinfo {title} {Effect of input noise dimension in {GANs}},\
		}\href@noop {} {\bibfield  {journal} {\bibinfo  {journal} {arXiv:2004.06882}\
			} (\bibinfo {year} {2020})}\BibitemShut {NoStop}%
		\bibitem [{\citenamefont {Ackley}\ \emph {et~al.}(1985)\citenamefont {Ackley},
			\citenamefont {Hinton},\ and\ \citenamefont {Sejnowski}}]{ackley1985RBM}%
		\BibitemOpen
		\bibfield  {author} {\bibinfo {author} {\bibfnamefont {D.~H.}\ \bibnamefont
				{Ackley}}, \bibinfo {author} {\bibfnamefont {G.~E.}\ \bibnamefont {Hinton}},\
			and\ \bibinfo {author} {\bibfnamefont {T.~J.}\ \bibnamefont {Sejnowski}},\
		}\bibfield  {title} {\bibinfo {title} {A learning algorithm for {Boltzmann}
				machines},\ }\href@noop {} {\bibfield  {journal} {\bibinfo  {journal}
				{Cognitive science}\ }\textbf {\bibinfo {volume} {9}},\ \bibinfo {pages}
			{147} (\bibinfo {year} {1985})}\BibitemShut {NoStop}%
		\bibitem [{\citenamefont {Alcazar}\ \emph {et~al.}(2020)\citenamefont
			{Alcazar}, \citenamefont {Leyton-Ortega},\ and\ \citenamefont
			{Perdomo-Ortiz}}]{Alcazar2020ClvsQuant}%
		\BibitemOpen
		\bibfield  {author} {\bibinfo {author} {\bibfnamefont {J.}~\bibnamefont
				{Alcazar}}, \bibinfo {author} {\bibfnamefont {V.}~\bibnamefont
				{Leyton-Ortega}},\ and\ \bibinfo {author} {\bibfnamefont {A.}~\bibnamefont
				{Perdomo-Ortiz}},\ }\bibfield  {title} {\bibinfo {title} {Classical versus
				quantum models in machine learning: insights from a finance application},\
		}\href {https://doi.org/10.1088/2632-2153/ab9009} {\bibfield  {journal}
			{\bibinfo  {journal} {Machine Learning: Science and Technology}\ }\textbf
			{\bibinfo {volume} {1}},\ \bibinfo {pages} {035003} (\bibinfo {year}
			{2020})}\BibitemShut {NoStop}%
		\bibitem [{\citenamefont {Hamilton}\ \emph
			{et~al.}(2019{\natexlab{a}})\citenamefont {Hamilton}, \citenamefont
			{Dumitrescu},\ and\ \citenamefont {Pooser}}]{Hamilton2018}%
		\BibitemOpen
		\bibfield  {author} {\bibinfo {author} {\bibfnamefont {K.~E.}\ \bibnamefont
				{Hamilton}}, \bibinfo {author} {\bibfnamefont {E.~F.}\ \bibnamefont
				{Dumitrescu}},\ and\ \bibinfo {author} {\bibfnamefont {R.~C.}\ \bibnamefont
				{Pooser}},\ }\bibfield  {title} {\bibinfo {title} {Generative model
				benchmarks for superconducting qubits},\ }\href@noop {} {\bibfield  {journal}
			{\bibinfo  {journal} {Physical Review A}\ }\textbf {\bibinfo {volume} {99}},\
			\bibinfo {pages} {062323} (\bibinfo {year} {2019}{\natexlab{a}})}\BibitemShut
		{NoStop}%
		\bibitem [{\citenamefont {Leyton-Ortega}\ \emph {et~al.}(2021)\citenamefont
			{Leyton-Ortega}, \citenamefont {Perdomo-Ortiz},\ and\ \citenamefont
			{Perdomo}}]{leyton2019robust}%
		\BibitemOpen
		\bibfield  {author} {\bibinfo {author} {\bibfnamefont {V.}~\bibnamefont
				{Leyton-Ortega}}, \bibinfo {author} {\bibfnamefont {A.}~\bibnamefont
				{Perdomo-Ortiz}},\ and\ \bibinfo {author} {\bibfnamefont {O.}~\bibnamefont
				{Perdomo}},\ }\bibfield  {title} {\bibinfo {title} {Robust implementation of
				generative modeling with parametrized quantum circuits},\ }\href@noop {}
		{\bibfield  {journal} {\bibinfo  {journal} {Quantum Machine Intelligence}\
			}\textbf {\bibinfo {volume} {3}},\ \bibinfo {pages} {1} (\bibinfo {year}
			{2021})}\BibitemShut {NoStop}%
		\bibitem [{\citenamefont {Coyle}\ \emph {et~al.}(2021)\citenamefont {Coyle},
			\citenamefont {Henderson}, \citenamefont {Le}, \citenamefont {Kumar},
			\citenamefont {Paini},\ and\ \citenamefont
			{Kashefi}}]{coyle2020generativeFinance}%
		\BibitemOpen
		\bibfield  {author} {\bibinfo {author} {\bibfnamefont {B.}~\bibnamefont
				{Coyle}}, \bibinfo {author} {\bibfnamefont {M.}~\bibnamefont {Henderson}},
			\bibinfo {author} {\bibfnamefont {J.~C.~J.}\ \bibnamefont {Le}}, \bibinfo
			{author} {\bibfnamefont {N.}~\bibnamefont {Kumar}}, \bibinfo {author}
			{\bibfnamefont {M.}~\bibnamefont {Paini}},\ and\ \bibinfo {author}
			{\bibfnamefont {E.}~\bibnamefont {Kashefi}},\ }\bibfield  {title} {\bibinfo
			{title} {Quantum versus classical generative modelling in finance},\
		}\href@noop {} {\bibfield  {journal} {\bibinfo  {journal} {Quantum Science
					and Technology}\ }\textbf {\bibinfo {volume} {6}},\ \bibinfo {pages} {024013}
			(\bibinfo {year} {2021})}\BibitemShut {NoStop}%
		\bibitem [{\citenamefont {Zhu}\ \emph {et~al.}(2019)\citenamefont {Zhu},
			\citenamefont {Linke}, \citenamefont {Benedetti}, \citenamefont {Landsman},
			\citenamefont {Nguyen}, \citenamefont {Alderete}, \citenamefont
			{Perdomo-Ortiz}, \citenamefont {Korda}, \citenamefont {Garfoot},
			\citenamefont {Brecque} \emph {et~al.}}]{Zhu2018}%
		\BibitemOpen
		\bibfield  {author} {\bibinfo {author} {\bibfnamefont {D.}~\bibnamefont
				{Zhu}}, \bibinfo {author} {\bibfnamefont {N.~M.}\ \bibnamefont {Linke}},
			\bibinfo {author} {\bibfnamefont {M.}~\bibnamefont {Benedetti}}, \bibinfo
			{author} {\bibfnamefont {K.~A.}\ \bibnamefont {Landsman}}, \bibinfo {author}
			{\bibfnamefont {N.~H.}\ \bibnamefont {Nguyen}}, \bibinfo {author}
			{\bibfnamefont {C.~H.}\ \bibnamefont {Alderete}}, \bibinfo {author}
			{\bibfnamefont {A.}~\bibnamefont {Perdomo-Ortiz}}, \bibinfo {author}
			{\bibfnamefont {N.}~\bibnamefont {Korda}}, \bibinfo {author} {\bibfnamefont
				{A.}~\bibnamefont {Garfoot}}, \bibinfo {author} {\bibfnamefont
				{C.}~\bibnamefont {Brecque}}, \emph {et~al.},\ }\bibfield  {title} {\bibinfo
			{title} {Training of quantum circuits on a hybrid quantum computer},\
		}\bibfield  {journal} {\bibinfo  {journal} {Science advances}\ }\textbf
		{\bibinfo {volume} {5}},\ \href {https://doi.org/10.1126/sciadv.aaw9918}
		{10.1126/sciadv.aaw9918} (\bibinfo {year} {2019})\BibitemShut {NoStop}%
		\bibitem [{\citenamefont {Hamilton}\ \emph
			{et~al.}(2019{\natexlab{b}})\citenamefont {Hamilton}, \citenamefont
			{Dumitrescu},\ and\ \citenamefont {Pooser}}]{hamilton2020hardwareQCBM}%
		\BibitemOpen
		\bibfield  {author} {\bibinfo {author} {\bibfnamefont {K.~E.}\ \bibnamefont
				{Hamilton}}, \bibinfo {author} {\bibfnamefont {E.~F.}\ \bibnamefont
				{Dumitrescu}},\ and\ \bibinfo {author} {\bibfnamefont {R.~C.}\ \bibnamefont
				{Pooser}},\ }\bibfield  {title} {\bibinfo {title} {Generative model
				benchmarks for superconducting qubits},\ }\href@noop {} {\bibfield  {journal}
			{\bibinfo  {journal} {Physical Review A}\ }\textbf {\bibinfo {volume} {99}},\
			\bibinfo {pages} {062323} (\bibinfo {year} {2019}{\natexlab{b}})}\BibitemShut
		{NoStop}%
		\bibitem [{\citenamefont {Gao}\ \emph {et~al.}(2021)\citenamefont {Gao},
			\citenamefont {Anschuetz}, \citenamefont {Wang}, \citenamefont {Cirac},\ and\
			\citenamefont {Lukin}}]{gao2021enhancing}%
		\BibitemOpen
		\bibfield  {author} {\bibinfo {author} {\bibfnamefont {X.}~\bibnamefont
				{Gao}}, \bibinfo {author} {\bibfnamefont {E.~R.}\ \bibnamefont {Anschuetz}},
			\bibinfo {author} {\bibfnamefont {S.-T.}\ \bibnamefont {Wang}}, \bibinfo
			{author} {\bibfnamefont {J.~I.}\ \bibnamefont {Cirac}},\ and\ \bibinfo
			{author} {\bibfnamefont {M.~D.}\ \bibnamefont {Lukin}},\ }\href@noop {}
		{\bibinfo {title} {Enhancing generative models via quantum correlations}}
		(\bibinfo {year} {2021}),\ \Eprint {https://arxiv.org/abs/2101.08354}
		{arXiv:2101.08354 [quant-ph]} \BibitemShut {NoStop}%
		\bibitem [{\citenamefont {Mcclean}\ \emph {et~al.}(2018)\citenamefont
			{Mcclean}, \citenamefont {Boixo}, \citenamefont {Smelyanskiy}, \citenamefont
			{Babbush},\ and\ \citenamefont {Neven}}]{Mcclear2018Barren}%
		\BibitemOpen
		\bibfield  {author} {\bibinfo {author} {\bibfnamefont {J.}~\bibnamefont
				{Mcclean}}, \bibinfo {author} {\bibfnamefont {S.}~\bibnamefont {Boixo}},
			\bibinfo {author} {\bibfnamefont {V.}~\bibnamefont {Smelyanskiy}}, \bibinfo
			{author} {\bibfnamefont {R.}~\bibnamefont {Babbush}},\ and\ \bibinfo {author}
			{\bibfnamefont {H.}~\bibnamefont {Neven}},\ }\bibfield  {title} {\bibinfo
			{title} {Barren plateaus in quantum neural network training landscapes},\
		}\href@noop {} {\bibfield  {journal} {\bibinfo  {journal} {Nature
					Communications}\ }\textbf {\bibinfo {volume} {9}} (\bibinfo {year}
			{2018})}\BibitemShut {NoStop}%
		\bibitem [{\citenamefont {Wright}\ \emph {et~al.}(2019)\citenamefont {Wright},
			\citenamefont {Beck}, \citenamefont {Debnath}, \citenamefont {Amini},
			\citenamefont {Nam}, \citenamefont {Grzesiak}, \citenamefont {Chen},
			\citenamefont {Pisenti}, \citenamefont {Chmielewski}, \citenamefont {Collins}
			\emph {et~al.}}]{Wright2019IonQ}%
		\BibitemOpen
		\bibfield  {author} {\bibinfo {author} {\bibfnamefont {K.}~\bibnamefont
				{Wright}}, \bibinfo {author} {\bibfnamefont {K.~M.}\ \bibnamefont {Beck}},
			\bibinfo {author} {\bibfnamefont {S.}~\bibnamefont {Debnath}}, \bibinfo
			{author} {\bibfnamefont {J.}~\bibnamefont {Amini}}, \bibinfo {author}
			{\bibfnamefont {Y.}~\bibnamefont {Nam}}, \bibinfo {author} {\bibfnamefont
				{N.}~\bibnamefont {Grzesiak}}, \bibinfo {author} {\bibfnamefont {J.-S.}\
				\bibnamefont {Chen}}, \bibinfo {author} {\bibfnamefont {N.}~\bibnamefont
				{Pisenti}}, \bibinfo {author} {\bibfnamefont {M.}~\bibnamefont
				{Chmielewski}}, \bibinfo {author} {\bibfnamefont {C.}~\bibnamefont
				{Collins}}, \emph {et~al.},\ }\bibfield  {title} {\bibinfo {title}
			{Benchmarking an 11-qubit quantum computer},\ }\href@noop {} {\bibfield
			{journal} {\bibinfo  {journal} {Nature communications}\ }\textbf {\bibinfo
				{volume} {10}},\ \bibinfo {pages} {1} (\bibinfo {year} {2019})}\BibitemShut
		{NoStop}%
		\bibitem [{\citenamefont {Sheng}\ and\ \citenamefont
			{Zhou}(2017)}]{sheng2017distributed}%
		\BibitemOpen
		\bibfield  {author} {\bibinfo {author} {\bibfnamefont {Y.-B.}\ \bibnamefont
				{Sheng}}\ and\ \bibinfo {author} {\bibfnamefont {L.}~\bibnamefont {Zhou}},\
		}\bibfield  {title} {\bibinfo {title} {Distributed secure quantum machine
				learning},\ }\href@noop {} {\bibfield  {journal} {\bibinfo  {journal}
				{Science Bulletin}\ }\textbf {\bibinfo {volume} {62}},\ \bibinfo {pages}
			{1025} (\bibinfo {year} {2017})}\BibitemShut {NoStop}%
		\bibitem [{\citenamefont {Song}\ \emph {et~al.}(2021)\citenamefont {Song},
			\citenamefont {Lim}, \citenamefont {Kwon}, \citenamefont {Adesso},
			\citenamefont {Wie{\'s}niak}, \citenamefont {Paw{\l}owski}, \citenamefont
			{Kim},\ and\ \citenamefont {Bang}}]{song2021secure}%
		\BibitemOpen
		\bibfield  {author} {\bibinfo {author} {\bibfnamefont {W.}~\bibnamefont
				{Song}}, \bibinfo {author} {\bibfnamefont {Y.}~\bibnamefont {Lim}}, \bibinfo
			{author} {\bibfnamefont {H.}~\bibnamefont {Kwon}}, \bibinfo {author}
			{\bibfnamefont {G.}~\bibnamefont {Adesso}}, \bibinfo {author} {\bibfnamefont
				{M.}~\bibnamefont {Wie{\'s}niak}}, \bibinfo {author} {\bibfnamefont
				{M.}~\bibnamefont {Paw{\l}owski}}, \bibinfo {author} {\bibfnamefont
				{J.}~\bibnamefont {Kim}},\ and\ \bibinfo {author} {\bibfnamefont
				{J.}~\bibnamefont {Bang}},\ }\bibfield  {title} {\bibinfo {title} {Quantum
				secure learning with classical samples},\ }\href@noop {} {\bibfield
			{journal} {\bibinfo  {journal} {Physical Review A}\ }\textbf {\bibinfo
				{volume} {103}},\ \bibinfo {pages} {042409} (\bibinfo {year}
			{2021})}\BibitemShut {NoStop}%
		\bibitem [{\citenamefont {Li}\ \emph {et~al.}(2021)\citenamefont {Li},
			\citenamefont {Lu},\ and\ \citenamefont {Deng}}]{li2021blind}%
		\BibitemOpen
		\bibfield  {author} {\bibinfo {author} {\bibfnamefont {W.}~\bibnamefont
				{Li}}, \bibinfo {author} {\bibfnamefont {S.}~\bibnamefont {Lu}},\ and\
			\bibinfo {author} {\bibfnamefont {D.-L.}\ \bibnamefont {Deng}},\ }\bibfield
		{title} {\bibinfo {title} {Quantum federated learning through blind quantum
				computing},\ }\href@noop {} {\bibfield  {journal} {\bibinfo  {journal}
				{Science China Physics, Mechanics \& Astronomy}\ }\textbf {\bibinfo {volume}
				{64}},\ \bibinfo {pages} {1} (\bibinfo {year} {2021})}\BibitemShut {NoStop}%
		\bibitem [{\citenamefont {Cheng}\ \emph {et~al.}(2017)\citenamefont {Cheng},
			\citenamefont {Chen},\ and\ \citenamefont {Wang}}]{Cheng2017TensorBorn}%
		\BibitemOpen
		\bibfield  {author} {\bibinfo {author} {\bibfnamefont {S.}~\bibnamefont
				{Cheng}}, \bibinfo {author} {\bibfnamefont {J.}~\bibnamefont {Chen}},\ and\
			\bibinfo {author} {\bibfnamefont {L.}~\bibnamefont {Wang}},\ }\bibfield
		{title} {\bibinfo {title} {Information perspective to probabilistic modeling:
				{Boltzmann} machines versus {Born} machines},\ }\href@noop {} {\bibfield
			{journal} {\bibinfo  {journal} {Entropy}\ }\textbf {\bibinfo {volume} {20}}
			(\bibinfo {year} {2017})}\BibitemShut {NoStop}%
		\bibitem [{\citenamefont {Alcazar}\ and\ \citenamefont
			{Perdomo-Ortiz}(2021)}]{alcazar2021enhancing}%
		\BibitemOpen
		\bibfield  {author} {\bibinfo {author} {\bibfnamefont {J.}~\bibnamefont
				{Alcazar}}\ and\ \bibinfo {author} {\bibfnamefont {A.}~\bibnamefont
				{Perdomo-Ortiz}},\ }\bibfield  {title} {\bibinfo {title} {Enhancing
				combinatorial optimization with quantum generative models},\ }\href
		{https://arxiv.org/abs/2101.06250} {\bibfield  {journal} {\bibinfo  {journal}
				{arXiv:2101.06250}\ } (\bibinfo {year} {2021})}\BibitemShut {NoStop}%
		\bibitem [{\citenamefont {Gili}\ \emph {et~al.}(2022)\citenamefont {Gili},
			\citenamefont {Mauri},\ and\ \citenamefont
			{Perdomo-Ortiz}}]{gili2022evaluating}%
		\BibitemOpen
		\bibfield  {author} {\bibinfo {author} {\bibfnamefont {K.}~\bibnamefont
				{Gili}}, \bibinfo {author} {\bibfnamefont {M.}~\bibnamefont {Mauri}},\ and\
			\bibinfo {author} {\bibfnamefont {A.}~\bibnamefont {Perdomo-Ortiz}},\
		}\bibfield  {title} {\bibinfo {title} {Evaluating generalization in classical
				and quantum generative models},\ }\href@noop {} {\bibfield  {journal}
			{\bibinfo  {journal} {arXiv preprint arXiv:2201.08770}\ } (\bibinfo {year}
			{2022})}\BibitemShut {NoStop}%
		\bibitem [{\citenamefont {Ioffe}\ and\ \citenamefont
			{Szegedy}(2015)}]{ioffe2015batchnorm}%
		\BibitemOpen
		\bibfield  {author} {\bibinfo {author} {\bibfnamefont {S.}~\bibnamefont
				{Ioffe}}\ and\ \bibinfo {author} {\bibfnamefont {C.}~\bibnamefont
				{Szegedy}},\ }\bibfield  {title} {\bibinfo {title} {Batch normalization:
				Accelerating deep network training by reducing internal covariate shift},\
		}in\ \href@noop {} {\emph {\bibinfo {booktitle} {International conference on
					machine learning}}}\ (\bibinfo {year} {2015})\ pp.\ \bibinfo {pages}
		{448--456}\BibitemShut {NoStop}%
		\bibitem [{\citenamefont {Xu}\ \emph {et~al.}(2015)\citenamefont {Xu},
			\citenamefont {Wang}, \citenamefont {Chen},\ and\ \citenamefont
			{Li}}]{xu2015ReLU}%
		\BibitemOpen
		\bibfield  {author} {\bibinfo {author} {\bibfnamefont {B.}~\bibnamefont
				{Xu}}, \bibinfo {author} {\bibfnamefont {N.}~\bibnamefont {Wang}}, \bibinfo
			{author} {\bibfnamefont {T.}~\bibnamefont {Chen}},\ and\ \bibinfo {author}
			{\bibfnamefont {M.}~\bibnamefont {Li}},\ }\bibfield  {title} {\bibinfo
			{title} {Empirical evaluation of rectified activations in convolutional
				network},\ }\href@noop {} {\bibfield  {journal} {\bibinfo  {journal}
				{arXiv:1505.00853}\ } (\bibinfo {year} {2015})}\BibitemShut {NoStop}%
		\bibitem [{\citenamefont {Spall}\ \emph {et~al.}(1992)\citenamefont {Spall}
			\emph {et~al.}}]{spall1992SPSA}%
		\BibitemOpen
		\bibfield  {author} {\bibinfo {author} {\bibfnamefont {J.~C.}\ \bibnamefont
				{Spall}} \emph {et~al.},\ }\bibfield  {title} {\bibinfo {title} {Multivariate
				stochastic approximation using a simultaneous perturbation gradient
				approximation},\ }\href@noop {} {\bibfield  {journal} {\bibinfo  {journal}
				{IEEE transactions on automatic control}\ }\textbf {\bibinfo {volume} {37}},\
			\bibinfo {pages} {332} (\bibinfo {year} {1992})}\BibitemShut {NoStop}%
		\bibitem [{\citenamefont {Salimans}\ \emph {et~al.}(2016)\citenamefont
			{Salimans}, \citenamefont {Goodfellow}, \citenamefont {Zaremba},
			\citenamefont {Cheung}, \citenamefont {Radford},\ and\ \citenamefont
			{Chen}}]{salimans2016improved}%
		\BibitemOpen
		\bibfield  {author} {\bibinfo {author} {\bibfnamefont {T.}~\bibnamefont
				{Salimans}}, \bibinfo {author} {\bibfnamefont {I.}~\bibnamefont
				{Goodfellow}}, \bibinfo {author} {\bibfnamefont {W.}~\bibnamefont {Zaremba}},
			\bibinfo {author} {\bibfnamefont {V.}~\bibnamefont {Cheung}}, \bibinfo
			{author} {\bibfnamefont {A.}~\bibnamefont {Radford}},\ and\ \bibinfo {author}
			{\bibfnamefont {X.}~\bibnamefont {Chen}},\ }\bibfield  {title} {\bibinfo
			{title} {Improved techniques for training {GAN}s},\ }\href
		{https://proceedings.neurips.cc/paper/2016/file/8a3363abe792db2d8761d6403605aeb7-Paper.pdf}
		{\bibfield  {journal} {\bibinfo  {journal} {Advances in neural information
					processing systems}\ }\textbf {\bibinfo {volume} {29}} (\bibinfo {year}
			{2016})}\BibitemShut {NoStop}%
		\bibitem [{\citenamefont {Borji}(2019)}]{borji2019GANmetrics}%
		\BibitemOpen
		\bibfield  {author} {\bibinfo {author} {\bibfnamefont {A.}~\bibnamefont
				{Borji}},\ }\bibfield  {title} {\bibinfo {title} {Pros and cons of {GAN}
				evaluation measures},\ }\href {https://doi.org/10.1016/j.cviu.2018.10.009}
		{\bibfield  {journal} {\bibinfo  {journal} {Computer Vision and Image
					Understanding}\ }\textbf {\bibinfo {volume} {179}},\ \bibinfo {pages} {41}
			(\bibinfo {year} {2019})}\BibitemShut {NoStop}%
		\bibitem [{\citenamefont {Szegedy}\ \emph {et~al.}(2016)\citenamefont
			{Szegedy}, \citenamefont {Vanhoucke}, \citenamefont {Ioffe}, \citenamefont
			{Shlens},\ and\ \citenamefont {Wojna}}]{Szegedy2016Inception}%
		\BibitemOpen
		\bibfield  {author} {\bibinfo {author} {\bibfnamefont {C.}~\bibnamefont
				{Szegedy}}, \bibinfo {author} {\bibfnamefont {V.}~\bibnamefont {Vanhoucke}},
			\bibinfo {author} {\bibfnamefont {S.}~\bibnamefont {Ioffe}}, \bibinfo
			{author} {\bibfnamefont {J.}~\bibnamefont {Shlens}},\ and\ \bibinfo {author}
			{\bibfnamefont {Z.}~\bibnamefont {Wojna}},\ }\bibfield  {title} {\bibinfo
			{title} {Rethinking the inception architecture for computer vision},\ }in\
		\href@noop {} {\emph {\bibinfo {booktitle} {Proceedings of the IEEE
					conference on computer vision and pattern recognition}}}\ (\bibinfo {year}
		{2016})\ pp.\ \bibinfo {pages} {2818--2826}\BibitemShut {NoStop}%
	\end{thebibliography}
	%

	\appendix
	\section{Details on the IonQ
		Hardware}\label{Apx:IonQ_device}
	The experimental circuits are implemented on an 11-qubit trapped ion processor based on $^{171}$Yb$^{+}$ ion qubits. The hyperfine levels of the ${}^2S_{1/2}$ ground state are used as the states of the qubit with $| 0 \rangle \equiv | F=0 , m_F = 0 \rangle $ and $| 1 \rangle \equiv |F=1 , m_F = 0 \rangle $. Measurement of the entire qubit register is achieved through state dependent fluorescence between $|1 \rangle $ and ${}^2 P_{1/2}$ states, with the scattered photons being collected through an aperture lens and passed through a dichroic mirror to an array of photon detectors. 
	
	The 11-qubit device is operated with automated loading of a linear chain of ions, which is then optically initialized with high fidelity. Computations are performed using a mode-locked $355$nm laser, which drives native single-qubit-gate (SQG) and two-qubit-gate (TQG) operations. TQG operations are done through the motional modes shared by all the ions, this allows for an  all to all connectivity topology. The native entangling operation, the M\o lmer S\o renson gate, written using Pauli operators is
	\begin{equation} \label{eq:ms_gate}
	\theta_{xx}^{i, j} = e^{-i \frac{\theta}{2} \sigma_x^i \sigma_x^j}.
	\end{equation}
	In order to maintain consistent gate performance, calibrations of the trapped ion processor are automated. Additionally, phase calibrations are performed for SQG and TQG sets, as required for implementing computations in queue and to ensure consistency of the gate performance.\\
	The device is commercially available through IonQ's cloud service. On the cloud, the system has all-to-all connectivity, an average 1-qubit gate fidelity of 99.35\%, an average 2-qubit gate fidelity of 96.02\% and SPAM fidelity of 99.3\%. For more details we refer to Ref.~\cite{Wright2019IonQ}.
	
	\section{The Quantum Circuit Born Machine}\label{Apx:QCBM_overview}
\begin{figure}
		\centering
		\includegraphics[width=0.9\linewidth]{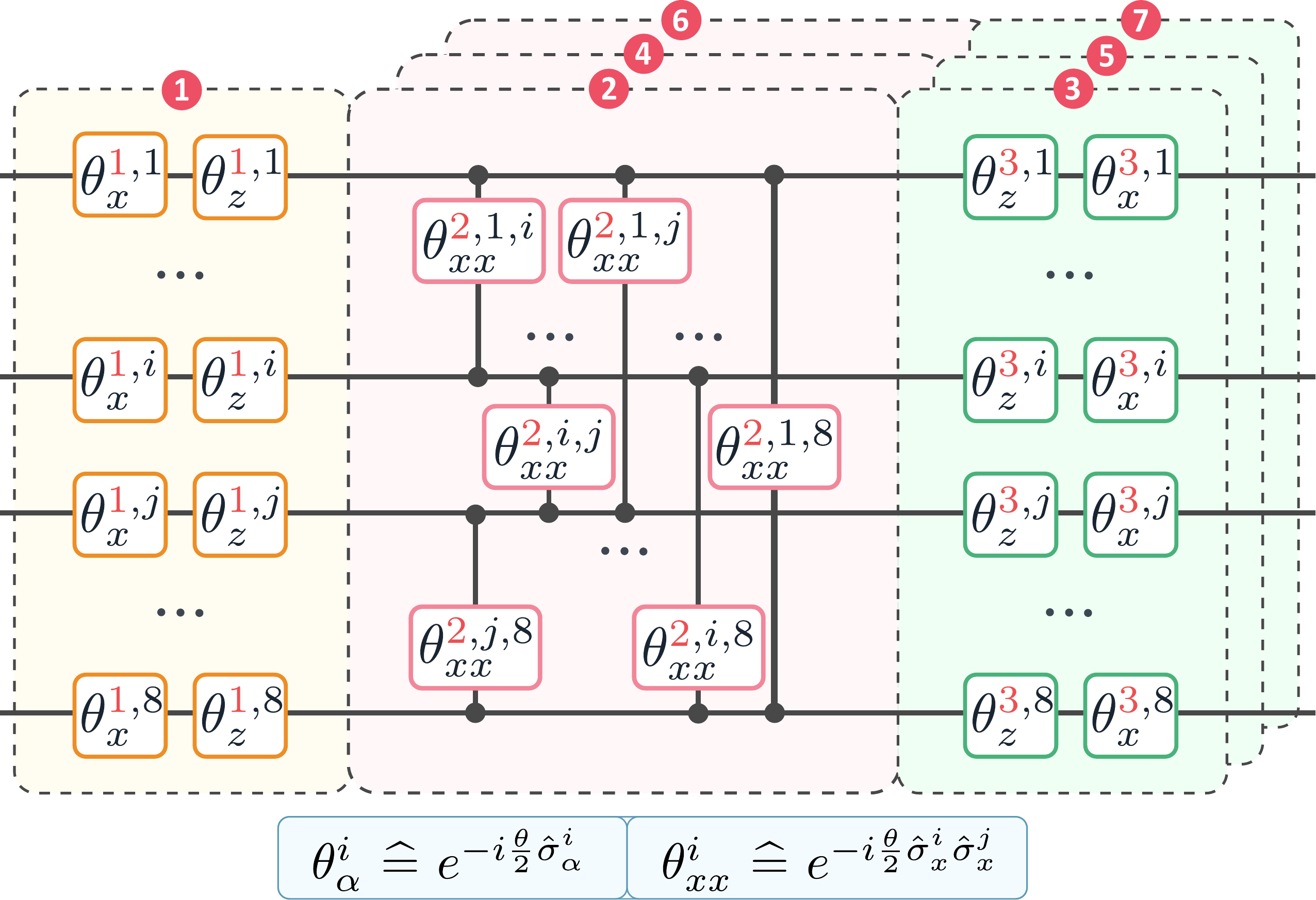}
		\caption{Quantum circuit ansatz for the QCBM. The ansatz is structured in layers to control expressivity of the model and fidelity of the prepared state. Throughout this work, we use a layer depth of 2 in order to maximize fidelity in the experimental implementation and enforce some interpolation between learned samples. The red numbers indicate the layer counting convention. For simulations, we used the all-to-all entangling layer as shown here, whereas for the experimental implementation, we instead adopted linear nearest-neighbor connectivity.}
		\label{fig:circuit_ansatz}
	\end{figure}

	Fig.~\ref{fig:circuit_ansatz} shows the quantum circuit ansatz used throughout this work to implement the QCBM state preparation unitary $U$ such that
	\begin{equation}\label{eq:qcbm_wavefunction_appendix}
	|\psi(\bm\theta)\rangle= \text{U}(\bm\theta)|0\rangle.
	\end{equation}
	The ansatz is inspired by capabilities of current ion-trap quantum devices and is structured in layers were expressivity of the model increases as layers are added. Although the QCBM equipped with this ansatz can become a powerful generative model, one needs to consider important trade-offs in the ansatz hyperparameter choice. For NISQ quantum devices, shallow quantum circuits are generally desired as deeper circuits can significantly decrease fidelity of the quantum states. Additionally, deep circuits oftentimes come with an excess number of parametrized quantum gates that enhance expressivity but can compromise trainability as well as creating a model that strongly overfits training data. For our work, we limit ourselves to 2 layers to minimize the number of gates used while introducing entanglement into the quantum state. In order to increase state fidelity for the experimental implementation, we additionally reduce the all-to-all connectivity of the $XX$ gates shown in Fig.~\ref{fig:circuit_ansatz} to a linear chain of entangling operations for the experiment on the IonQ quantum device. The circuit parameters are initialized with a warm start such that the QCBM encodes a uniform distribution in computational basis as well as the $o/t$ bases discussed in the main text and Appendix~\ref{Apx:multi_basis}.\\\\
	In our application, the QCBM is trained by minimizing the \textit{clipped negative log-likelihood}
	\begin{equation}\label{eq:clipped_nll}
	\mathcal{L}(\bm\theta)=-\sum_x p(\textbf{x})\log \text{max}(q_{\bm\theta}(\textbf{x}),\epsilon)
	\end{equation}
	where $p(\textbf{x})$ is the probability over all training data samples $\textbf{x}$ and $q_{\bm\theta}(\textbf{x}) = |\langle \textbf{x}|\psi(\bm\theta)\rangle|^2$ is the QCBM model distribution. Minimizing this function is equivalent to maximizing the expression in Eq.~\ref{eq:cost_prior}.  A regularization constant $\epsilon$ prevents singularity of the logarithm for samples with zero probability. The probability distribution $q_{\bm\theta}(\textbf{x})$ of samples \textbf{x} is estimated by sampling the prepared state and accumulating the measurements.
	\section{The Multi-Basis Technique}\label{Apx:multi_basis}
	In this work, we introduce a multi-basis technique for quantum circuit-based models to expand the repertoire of quantum machine learning researchers.
	
	Commonly, when referring to \textit{sampling} a generative model, one means generating instances of data that follow the encoded probability distribution. For classical models, one is limited to one basis - the computational basis. In quantum models, this is not the case. When encoding a probability distribution into a qubit wavefunction, as is the case with a Quantum Circuit Born Machine (QCBM), the learned wavefunction contains a potential family of sample distributions which are accessible by measuring in different bases. These additional distributions, or more specifically, projections of the wavefunction, can be evaluated by applying arbitrary post-rotations to the quantum registers before measurement. In this work, we explore the questions of whether we can enhance a generative model by including measurements in additional bases and how we can maximize the benefit of measuring additional basis sets. For this purpose, we quantitatively compare performance inside the QC-AAN framework using samples from an orthogonal measurement basis, i.e., the Y-basis, and flexible bases where the post-rotation angles are trained together with the ansatz parameters. This is done by doubling the latent space in the discriminator and training the samples of the multi-basis QCBM on the respective latent activations. While we do not explore a variety of possible basis choices, we argue that the trainable multi-basis model should converge to a learned constellation of measurement bases which is close to optimal for this specific training instance. Future work will need to study a wider range of variations of the multi-basis technique to uncover how to most efficiently utilize near-term quantum devices.
	
	As an explicit example of the sample concatenation, assume samples $\textbf{s} = \text{1010}$ and $\textbf{s}_{o/t} = \text{1100}$ which samples in computational and $o/t$ basis, respectively. Together, they then define $\textbf{s*} = \text{10101100}$. For a series of measurements in the computational and $o/t$ bases, the assignment of which pair of measurements is forwarded to the neural network is arbitrary as there is no direct correlation between the computational basis and $o/t$ basis distributions other than that they obey the normalization constraint of the QCBM wavefunction. This technique generalizes to the measurement and concatenation of samples of any observable able to be measured on a quantum device.
	
	A more subtle advantage of this multi-basis technique in the context of generative modelling with a QCBM is that the prior size $N$ (see Sec.~\ref{sec:GANs}) is doubled from $n$ to $2n$ where $n$ is the number of qubits. The effective sample space for the binary samples of the QCBM scales like $2^N$ which thus increases from $2^n$ to $2^{2n}$. Although the sample $\textbf{s*}$ lives in the $\{0,1\}^{2n}$ space, the multi-basis QCBM does not have access to the full $2n$ qubit space because of the normalization constraint of the wavefunction. However, increasing the sample dimension and additionally the amount of information encoded and utilized from small near-term quantum computers, here proved of immense value by enhancing the expressivity and the robustness of the hybrid QC-AAN model considered. 
	\section{Associative Adversarial Networks}
	Associative Adversarial Networks (AANs) were first proposed in Ref.~\cite{arici2016associative} as an extension of the popular Generative Adversarial Networks (GANs). Their purpose was to improve trainability and consequently general performance of GANs by modeling and reparametrizing the generator $G$'s prior distribution with a restricted Boltzmann machine (RBM). The objective functional for the AAN can be written as
	\begin{equation}\label{eq:cost_AAN}
	\mathcal{C}_{AAN} = \mathcal{C}_{GAN} \circ \mathcal{C}_{q},
	\end{equation}
	and consists of the traditional mini-max game objective $\mathcal{C}_{GAN}$ of a GAN in Eq.~\ref{eq:GAN_costfunction}, as well as the prior loss $\mathcal{C}_{q}$ in Eq.~\ref{eq:cost_prior} which is optimized in tandem with the GAN. Optimizing $\mathcal{C}_{q}$ maximizes the likelihood of the RBM distribution
	\begin{equation}
	q(\textbf{\textbf{z}}) = \sum_\textbf{h}\frac{e^{-E(\textbf{z},\textbf{h})}}{Z},
	\end{equation}
	given the latent distribution $p_l$ in the discriminator $D$. $E(\textbf{z},\textbf{h})$ is the energy functional for the RBM with visible units $\textbf{z}$ and hidden units $\textbf{h}$, and $Z=\sum_{\textbf{z},\textbf{h}}e^{-E(\textbf{z},\textbf{h})}$ is the partition function. For more details, we refer the reader to the original AAN paper in Ref.~\cite{arici2016associative}.
	\section{Simulated QC-AAN results for 6 and 8 qubits}\label{Apx:simulated_results}
\begin{figure}
		\centering
		\includegraphics[width=1.0\linewidth]{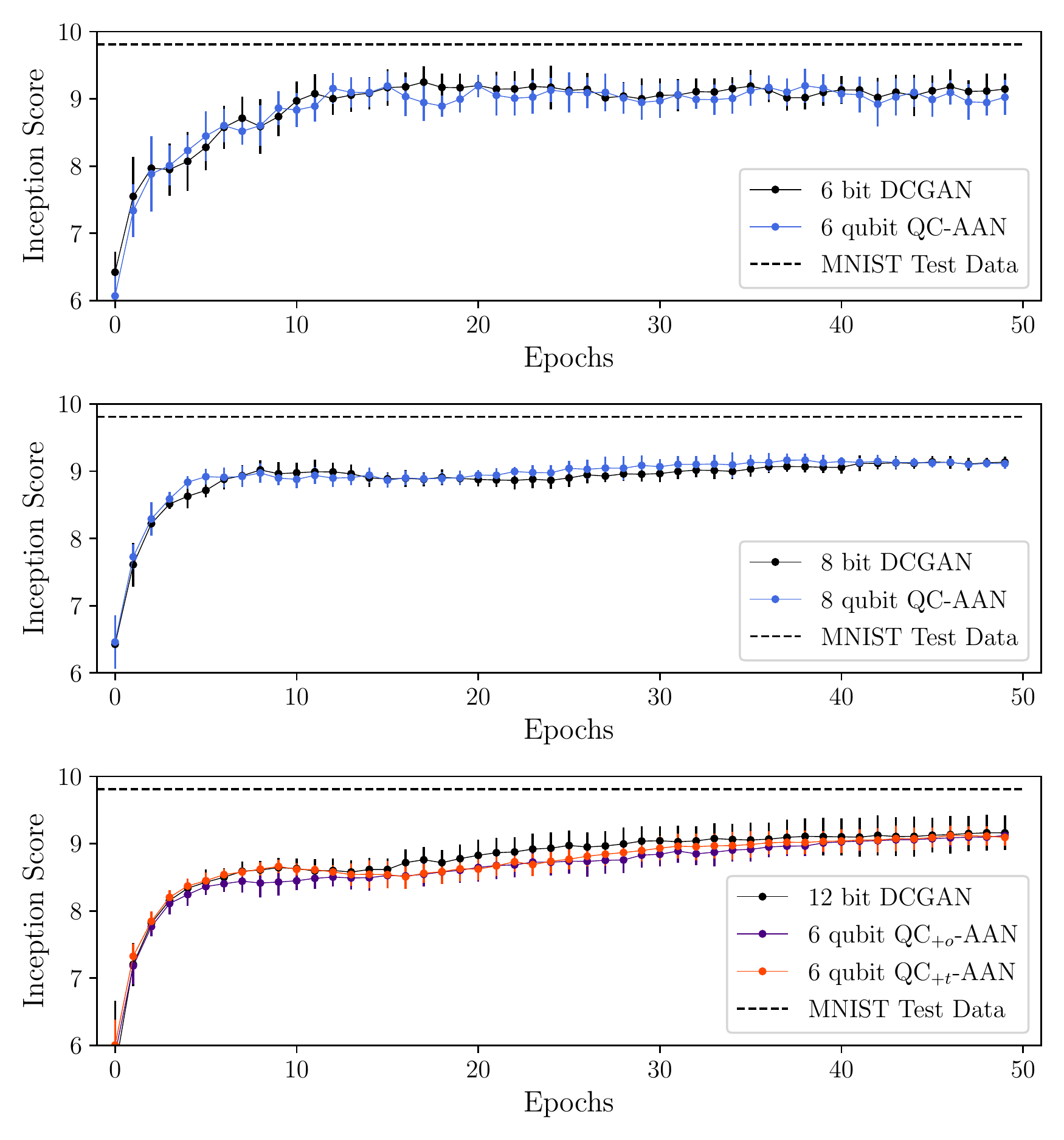}
		\caption{Simulation results of QC-AANs with 6- and 8 qubits relative to comparable DCGANs with uniform prior distribution. Depicted are average Inception Scores and standard deviations of 10 independent training repetitions per model. We observe no significant improvement for the 6- and 8 qubit QC-AANs, as well as for the QC$_{+o/t}$-AANs with 6 qubits. The first improvements were observed for the 8 qubit QC$_{+o/t}$-AAN, for which we refer to Figure~\ref{fig:IonQ_full_run_images} in the main text.}
		\label{fig:all_I_scores}
	\end{figure}

	To benchmark the performance of a QC-AAN with few qubits, we compare average Inception Scores (IS) of our QC-AANs, with 6- and 8-qubit multi-basis QCBM in the model prior. For this work, we only compare models with the same prior dimension to isolate the effect that re-parametrization of the prior distribution has on GAN training. In Fig.~\ref{fig:all_I_scores} we show that for 6- and 8-qubit QC-AANs, we could not achieve an advantage in learning a non-trivial prior. For a 6 (8) qubit QCBM, there are only 64 (256) distinct samples available for the neural network to map to high-quality images. Since the IS is very sensitive to class imbalance of the generated images, we see that modeling those priors does not lead to meaningful improvements. In fact, the results shown in Fig.~\ref{fig:all_I_scores} for the QC-AAN with 6- and 8 qubits were obtained by only minimally disturbing the uniform distribution.
	The QC$_{+o/t}$-AANs with the multi-basis technique show interesting results where for 6 qubits, we almost reach the performance of a 12 bit DCGAN, whereas with 8 qubits, we outperform a 16 bit DCGAN (see Fig.~\ref{fig:IonQ_full_run_images} in main text). This is despite the fact that the QC$_{+o/t}$-AAN models are restricted to an effictive sub-space compared to a Hilbert space with double the number of qubits.\\ 
	Note, that these results are specific to our neural network architecture. It is possible that more expressive and computationally intensive neural networks face less challenges in learning a great model with a uniform prior distribution. For a general learning task and network architecture, the QC-AAN algorithm allows to consider the hyperparameters of the trainable prior as hyperparameters towards a more successful overall generative algorithm.\\
	
	The 8 qubit QC$_{+o/t}$-AANs operate on a space with $2^{16} =  65,536$ potential samples. Although the multi-basis QCBM is a binary model, with this amount of different images, an Inception Score of occasionally over $9.5$ can be considered to be comparable to state-of-the-art GANs with continuous priors. In fact, Ref.~\cite{brock2019BigGAN} argues that binary units may perform at least as good as continuous uniform or normal distributions.
	\section{Neural Network Architectures and Training}\label{Apx:network_achitecture}
	\begin{figure}
		\centering
		\textbf{Generator $G$}
		\includegraphics[width=0.9\linewidth]{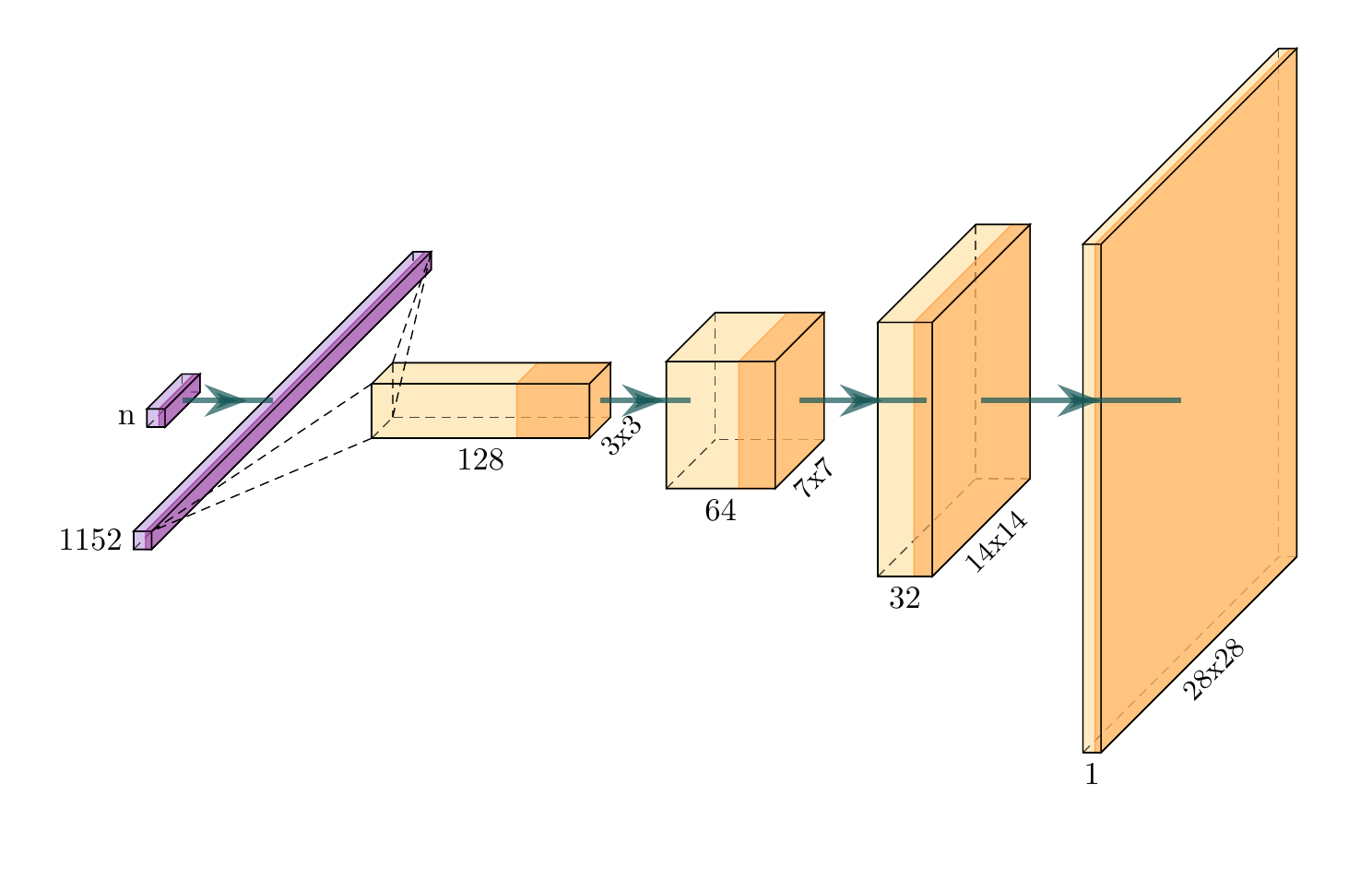}\\
		\textbf{Discriminator $D$}
		\includegraphics[width=0.9\linewidth]{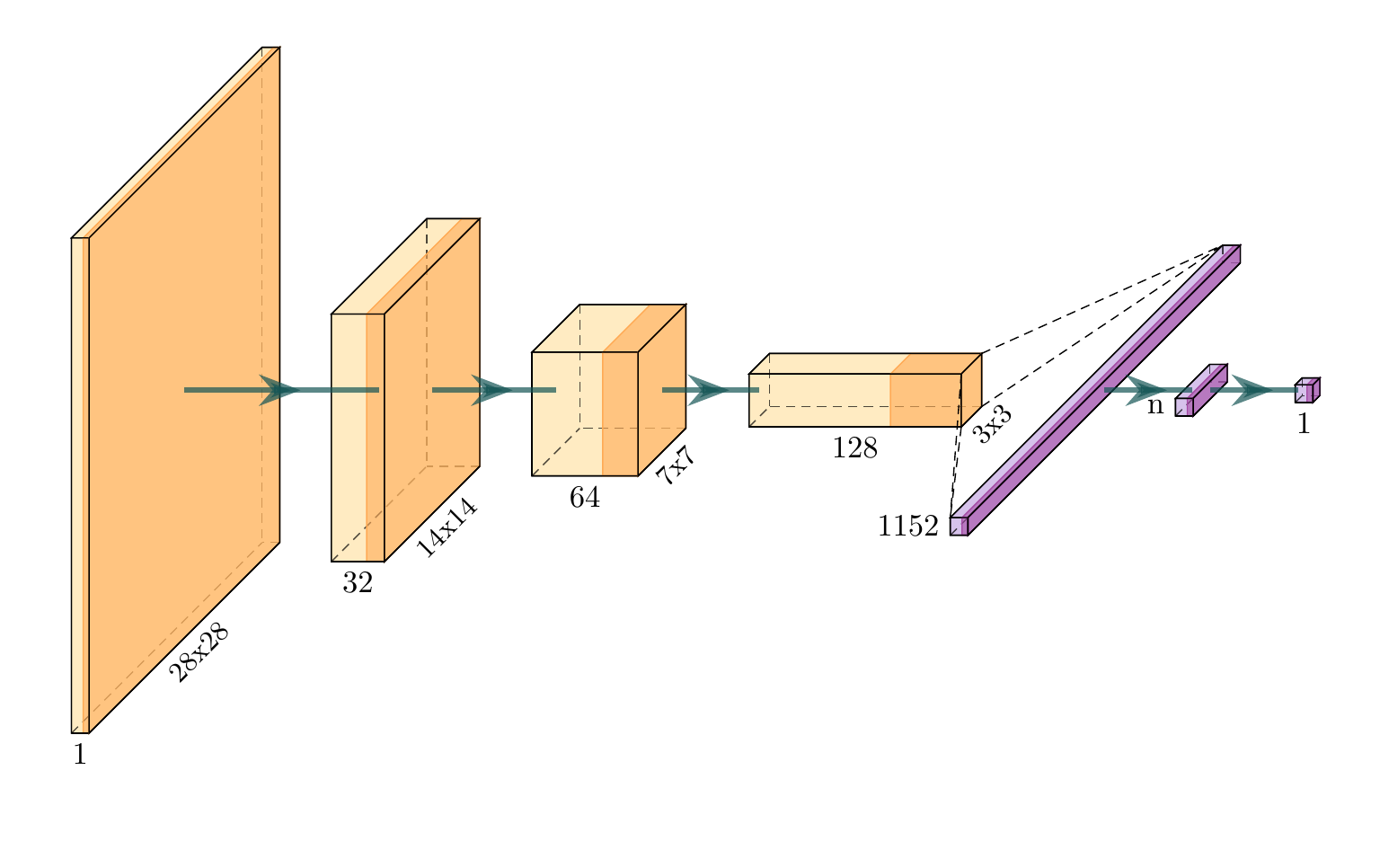}
		\caption{Schematic neural network architectures of the generator $G$ and discriminator $D$ used throughout this work for the MNIST data set of handwritten Digits. Purple color indicate a 1d layer of nodes whereas the orange blocks represent 2d convolutional layers with $128$, $64$ and $32$ channels. The dark color change inside each block indicates the application of a non-linear activation function after every linear transformation. $G$ and $D$ are approximately inverse, although this is not strictly required. Note that the second to last layer in $D$ represents the latent space which contains $n$ bits, the same size chosen for the prior of $G$.}
		\label{fig:network_architecture}
	\end{figure}
	Fig.~\ref{fig:network_architecture} shows a schematic overview of the network architectures of the generator $G$ and discriminator $D$ neural networks used throughout this work. We emphasize that these architectures are employed for both our hybrid QC-AANs, and the classical DCGANs. The networks have approximately inverse structure with three convolutional layers, although it is not generally required for stable GAN training. The second to last layer in $D$ (latent space) has the same size as the first layer in $G$ (prior space) to be able to train the quantum model in the QC-AAN on the latent activations of $D$. The total number of parameters in each network amounts to approximately $2.77\times 10^5$. The multi-basis QCBM in the QC-AAN adds between $31$ and $52$ trainable parameters for the 8 qubit model, equating to an increase of $0.02\%$ in the total number of parameters.
	
	All convolutional layers have batch normalization~\cite{ioffe2015batchnorm} and \textit{leaky rectified linear unit} (ReLU) activation functions~\cite{xu2015ReLU}. In training $D$, a small percentage ($3\%$) of training samples have their label flipped and label smoothing is applied to the training images. The optimizer for both networks is the ADAM optimizer with parameters $[\beta_1,\beta_2] = [0.5,0.9]$. Generally, our neural network architectures and hyperparameter choices are inspired by successful practices in modern DCGANs~\cite{radford2016dcgan}.
	
	\section{Training Details of QCBM as Model Prior}\label{Apx:qcbm_training}
	The QCBM in this work implements a hardware efficient ansatz inspired by capabilities of ion-trap quantum computers (see Fig.~\ref{fig:circuit_ansatz}). The layer depth of the ansatz is chosen to be shallow with only one layer of single-qubit and entangling gates respectively. For the numerical simulations, we chose an all-to-all connectivity between qubits, whereas in the hardware implementation, we used linear connectivity to improve the state fidelity. For the case of the MNIST training set, we did not observe on average significant negative effects in reducing the ansatz connectivity. We expect that for more challenging generative modeling tasks, the circuit ansatz will play a more crucial role.
	
	Over one QC-AAN training epoch on the entire MNIST data set with $N=60,000$ images, we perform an update of the QCBM parameters every $100$ batches for the simulation and every $600$ batches for the experimental implementation. The latter implies one training step per training epoch. We use the Simultaneous Perturbation Stochastic Approximation (SPSA) algorithm~\cite{spall1992SPSA} for training the parametrized quantum circuit to adapt the model distribution of the QCBM while minimizing calls of the quantum device. The gradients are evaluated with 1000 readout measurements (shots). For the experimental implementation on the quantum device, we sample the 8 qubit distributions with $10^4$ shots per measured basis and are able to construct multi-basis samples appropriately by resampling those measurements until the next training step. For the numerical simulations, this resampling was not performed and circuits were evaluated with as many shots as required for the GAN, i.e. for each image generated by the generator.\\
	One technique that has shown to stabilize training for the QC-AAN is to freeze the prior, i.e. to fix the QCBM parameters, after a certain number of training epochs. Altering the prior distribution significantly in the latter stages of training has shown to destabilize training and lead to visibly worse images. This is likely due to the converging generator network where parameters are slowly settling. Altering the prior at these stages introduces a new impulse for training, which is not matched by the decreasing step sizes of the ADAM optimizer of the networks (see Appendix~\ref{Apx:network_achitecture}). Throughout this work, we freeze the prior after 10 epochs.
	
	\section{QCBM and RBM in the AAN-Framework}\label{Apx:rbm_qcbm}
\begin{figure}[t]
		\centering
		\includegraphics[width=0.99\linewidth]{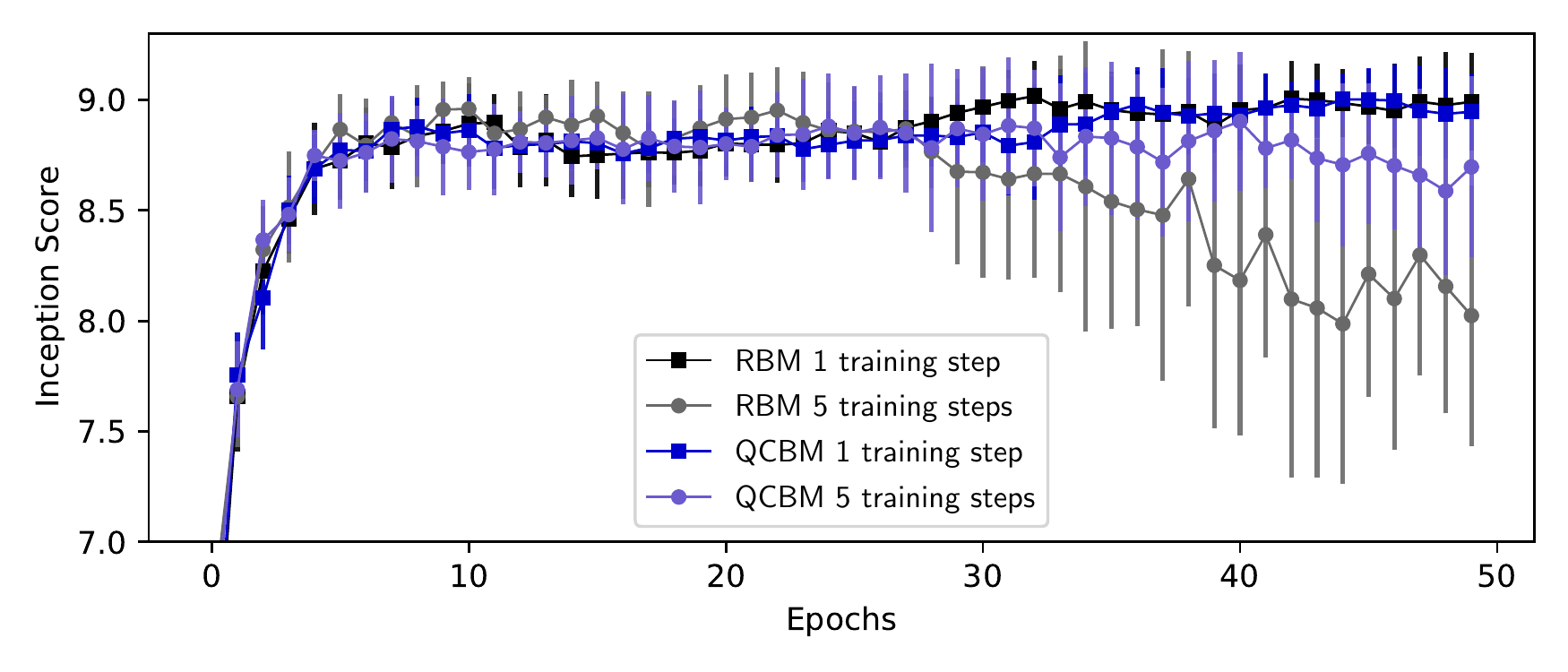}
		\includegraphics[width=0.49\linewidth]{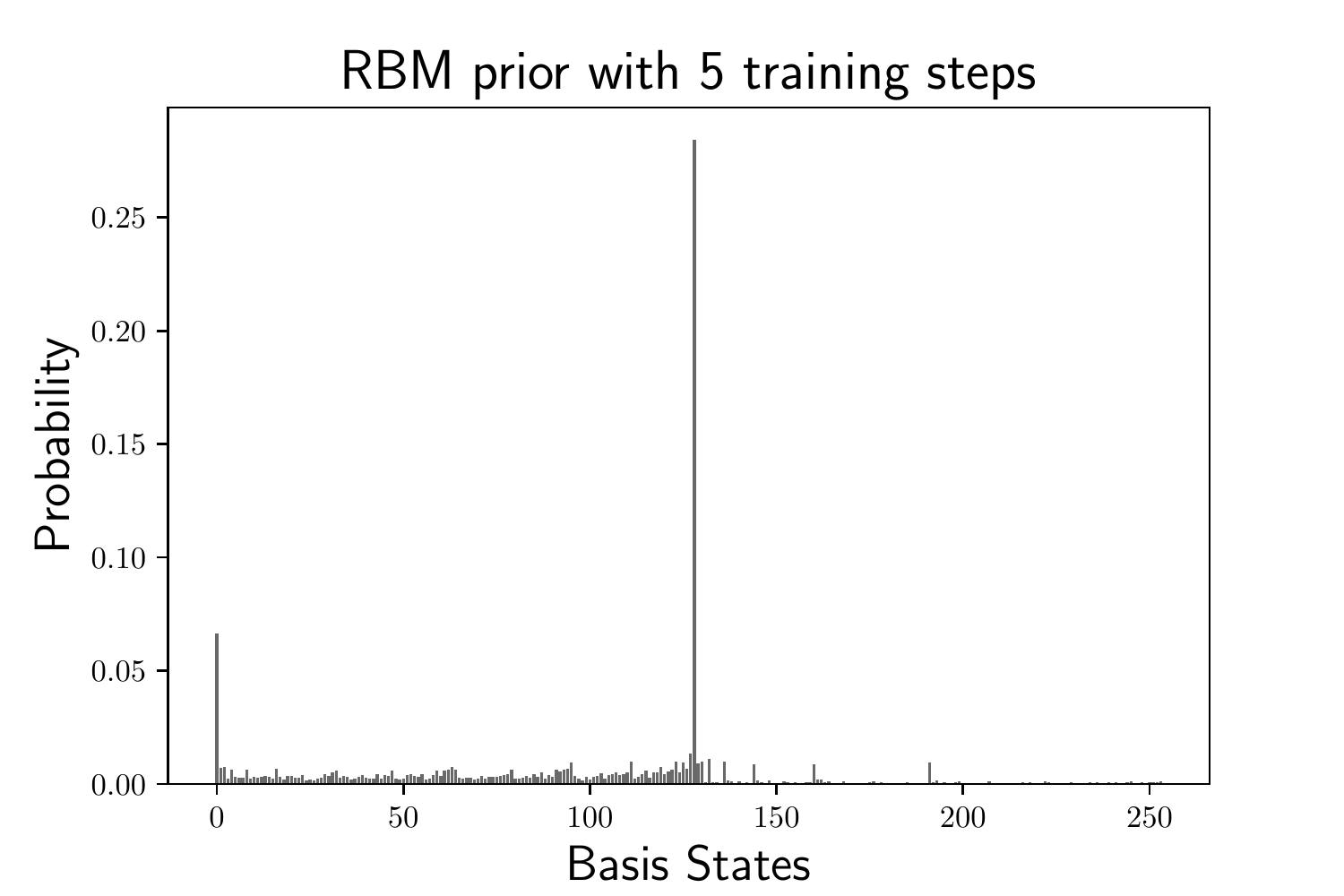}
		\includegraphics[width=0.49\linewidth]{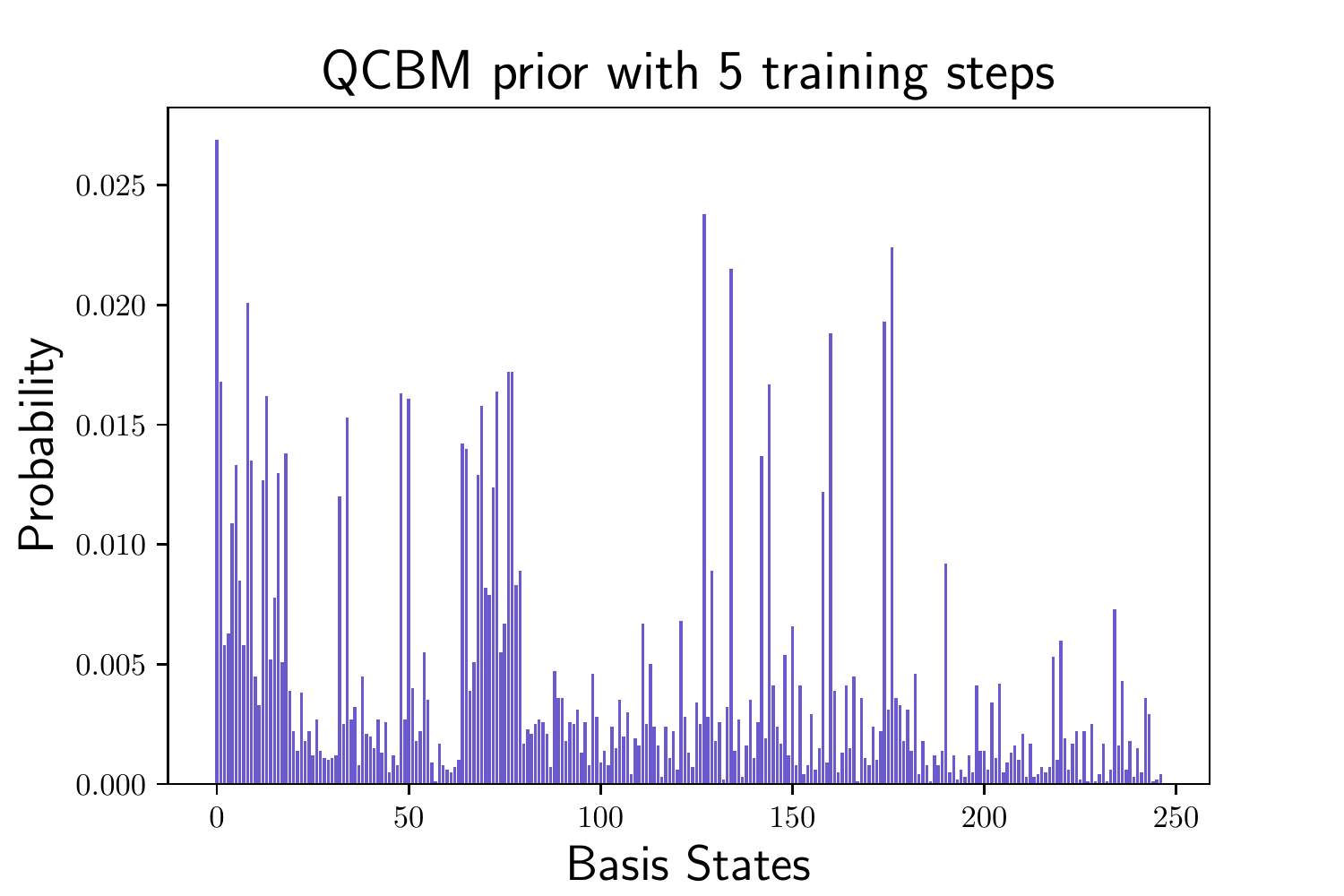}
		\caption{Example comparison of two hyperparameter configurations for an AAN and QC-AAN, which implement a RBM and QCBM as the model prior, respectively. A training protocol of 5 training steps per training instance appears to be too much variation in the GAN prior for stable training, but unlike a QCBM, a RBM is prone to becoming unstable under sub-optimal configurations and sampling only one distribution mode. When that happens, training of the algorithm fails.}
		\label{fig:RBM_QCBM_comparison}
	\end{figure}

	Quantum Circuit Born Machines (QCBMs) are promising quantum generative models that offer global sampling capabilities of the encoded distribution and additionally provide access to quantum measurements that may be beneficial in learning a strong model. One of those properties, which we leverage in this work, is the possibility to measure in additional bases. Still, one needs to weigh the costs and benefits for such a quantum model when Restricted Boltzmann Machines (RBMs) are light-weight generative models with efficient but local sampling algorithm. Ref.~\cite{Anschuetz2019QAAN} shows that the AAN framework with an RBM modeling the Generator's prior could be improved by instead implementing a Quantum Boltzmann Machine. Finding the best hyperparameters for their respective models is a notoriously difficult task and comparing models across all possible hyperparameter combinations is in general unfeasible. In this work, we argue that, given a poor choice of hyperparameters and the same number of model parameters, RBMs can easily become unstable in our smooth learning protocol while QCBMs in the QC-AAN retain their stability. Fig.~\ref{fig:RBM_QCBM_comparison} shows average training performances of 10 individual QC-AANs and AANs, both with 8-dimensional prior on the MNIST training set. The priors are trained in a smooth transition protocol where every 100 batches, their distributions are altered by 1 or 5 training steps of stochastic gradient descent with a step size of $0.01$. A priori, this step size seems small enough that both of the hyperparemter options of 1 or 5 training steps do not seem unreasonable. However, Fig.~\ref{fig:RBM_QCBM_comparison} shows that 5 training steps is in fact a sub-optimal hyperparameter choice. Notably, the RBM suffers from its native local sampling technique and can become very unstable. We also provide the sampled prior distributions of the worst performing models across 10 training runs. It is apparent that the RBM has reached a distribution which it cannot effectively sample. If this happens at any stage of our smooth transition training protocol, further training of the AAN becomes impossible. For the QCBM, the change in performance between two hyperparameter choices is more continuous. This example indicates a potentially improved robustness of the QC-AAN framework as compared to the AAN framework.
	
	\section{Inception Score Definition \& Discussion}\label{Apx:IS}
	The \textit{Inception Score} (IS)
	\begin{equation}\label{Inception_score}
	\begin{aligned}
	\text{IS}(G)  = \exp \left( \mathbb{E}_{\Tilde{x}\sim G}\big[\text{KL}\left[ p (y | \Tilde{x})|| p (y)\right]\big]\right)
	\end{aligned}
	\end{equation}
	is a popular metric for evaluating GANs. For a given generator \textit{G}, it measures the quality $p(y|\Tilde{x})$ of generated images $\Tilde{x}$ and also their diversity $p(y)$ across all possible classes $y$ of the original data set. Quantitatively, this is done by calculating the KL divergence between the posterior probability distribution $p(y|\Tilde{x})$ and the prior probabilities $p(y)$ for each class label. For details, we refer to Ref.~\cite{salimans2016improved}.
	The IS is a human-readable metric with values between $1$ and the number of total classes in the data set, i.e. $10$ for the MNIST data set of handwritten digits. Although it has been proven to be very useful.~\cite{salimans2016improved}, one of the main criticisms of the IS is that it does not depict the realism of generated images for a human observer because it is calculated with classifiers which are trained to search and find exactly the class labels that they have been trained for. Images can either be warped or noisy and still achieve very high classification certainty~\cite{borji2019GANmetrics}. In this work specifically, we achieve a surprisingly high IS with models that implement 6- and 8-dimensional priors, resulting in only 64 and 256 distinct images in total. Although the IS is high for those models, a human observer does not judge them as being particularly clear images. For such limited models, there arises an interesting effect where the discriminator can remember all images generated by the generator, constantly pushing it away, preventing further convergence and thus introducing noisy artifacts. Another classifier will clearly identify the digits as member of their particular class, regardless of the noise. Nevertheless, IS is a straight-forward quantitative performance measure for GANs that commonly correlates well with human perception.\\\\
	The IS is commonly calculated with use of the pre-trained Inception-v3 Network~\cite{Szegedy2016Inception} as a proxy to calculate the probabilities in Eq. \ref{Inception_score}. Since the Inception-v3 Networks can only be applied to colored data, we instead utilize a convolutional classifier with approximately $99.3\%$ accuracy on the MNIST data set to calculate Inception Scores.
	\section{Practical quantum advantage}\label{Apx:practical}
\begin{figure}[ht]
		\centering
		\includegraphics[width=0.9\linewidth]{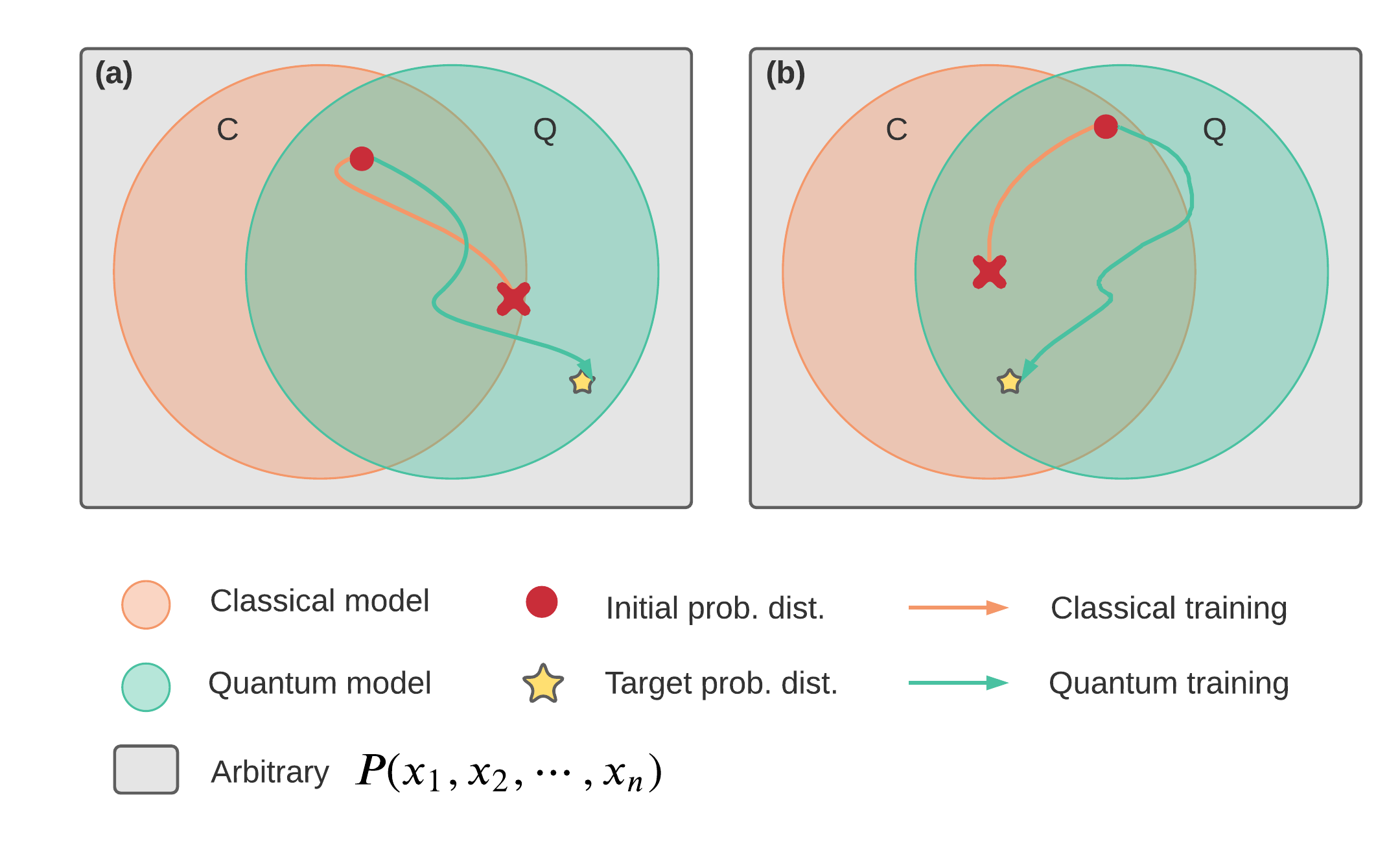}
		\caption{
			Sketch for two potential scenarios for practical quantum advantage in generative modeling. The large overlapping circles indicate the set of probability distributions which could be expressed within two given parametrized classical or quantum generative models. Note that this does not represent all distributions which could possibly be expressed by any classical or quantum model. \textbf{a}) The target distribution lies outside of the space that can be expressed by the classical model, similar to Fig.~5 in Ref.~\cite{Coyle2019}. This might be due to the distribution not being tractable classically, or because this specific parametrized model cannot reach it. \textbf{b}) The target distribution lies in a region which both models can reach. However, both models do not traverse the space equally and the classical model might be hindered, for example, by the presence of more local minima or due to pitfalls in conventional heuristics. As indicated in Fig.~\ref{fig:RBM_QCBM_comparison}, training success could be significantly different for the classical and quantum algorithms.
		}
		\label{fig:practical_quantum_advantage}
	\end{figure}

	Although theoretical gaps between classical and quantum generative modeling have been provided (see e.g. Refs.~\cite{sweke2020learnability, gao2021enhancing, Coyle2019, hinsche2021learnability}), these are not  sufficient to achieve a \textit{practical quantum advantage} since such a gap is not ensured to manifest in real-world data sets. Also, it is not evident that quantum algorithms can only outperform classical methods in cases where a theoretical gap exists for the desired target distribution. For example, the results presented in Ref.~\cite{Coyle2019} imply that for a certain family of classically intractable target distributions, also model distributions during training may be intractable. As schematically illustrated in Fig.~\ref{fig:practical_quantum_advantage}b, quantum resources offer a different tool set for tackling problems such that quantum algorithms could exhibit a practical quantum advantage by other means, even in cases where the generative task is within reach of the classical parametrized model, as is the case for the GANs considered in this work. For example, it is oftentimes overlooked that the training of an algorithm, here specifically a generative algorithm, is essential to its final performance and is arguably more important than tractability of the final outcome. If a quantum model provides more stable training or can navigate the cost function landscape more effectively, as indicated in Appendix~\ref{Apx:rbm_qcbm} and Fig.~\ref{fig:RBM_QCBM_comparison}, this opens up another path towards a practical quantum advantage.\\\\
	
\end{document}